\documentstyle[epsfig]{lamuphys}
\makeatletter
\let\chapter\hid@chapter
\makeatother
\renewcommand{\arraystretch}{1.3}
\newcommand{\cL}{{\cal L}}
\newcommand{\dslash}{\mbox{$\partial\hspace{-.55em}/$}}
\newcommand{\Dslash}{\mbox{${D\hspace{-.65em}/}$}}
\begin{document}
\pagenumbering{arabic}
\thispagestyle{empty}
\setcounter{page}{0}
\begin{flushright}
UWThPh-1998-18\\
May 1998
\end{flushright}

\vfill

\begin{center}
{\LARGE \bf CHIRAL SYMMETRY*} \\[40pt]

\large
Gerhard Ecker\\[0.5cm]
Institut f\"ur Theoretische Physik, Universit\"at Wien\\
Boltzmanngasse 5, A--1090 Wien, Austria \\[10pt]
\end{center}

\vfill

\begin{abstract}
Broken chiral symmetry has become the basis for a unified
treatment of hadronic interactions at low energies. After
reviewing mechanisms for spontaneous chiral symmetry breaking,
I outline the construction of the low--energy effective field
theory of the Standard Model called chiral perturbation theory.
The loop expansion and the renormalization procedure for this
nonrenormalizable quantum field theory are developed. Evidence 
for the standard scenario with a large quark condensate is
presented, in particular from high--statistics lattice
calculations of the meson mass spectrum. Elastic pion--pion
scattering is discussed as an example of a complete calculation
to $O(p^6)$ in the low--energy expansion. The meson--baryon
system is the subject of the last lecture. After a short summary
of heavy baryon chiral perturbation theory, a recent analysis of
pion--nucleon scattering to $O(p^3)$ is reviewed. Finally, I
describe some very recent progress in the chiral approach to the
nucleon--nucleon interaction.
\end{abstract}

\vfill

\begin{center}
Lectures given at the \\[5pt]
37. Internationale Universit\"atswochen f\"ur Kern-- und 
Teilchenphysik \\[5pt] 
Schladming, Austria, Feb. 28 - March 7, 1998 \\[5pt]
To appear in the Proceedings
\end{center}

\vfill

\noindent *  Work supported in part by TMR, EC--Contract No. ERBFMRX--CT980169
(EURODA$\Phi$NE). 

\newpage

\title{Chiral Symmetry}

\author{Gerhard\,Ecker}

\institute{Inst. Theor. Physik, Universit\"at Wien, Boltzmanng. 5,
A-1090 Wien, Austria}

\maketitle

\begin{abstract}
Broken chiral symmetry has become the basis for a unified
treatment of hadronic interactions at low energies. After
reviewing mechanisms for spontaneous chiral symmetry breaking,
I outline the construction of the low--energy effective field
theory of the Standard Model called chiral perturbation theory.
The loop expansion and the renormalization procedure for this
nonrenormalizable quantum field theory are developed. Evidence 
for the standard scenario with a large quark condensate is
presented, in particular from high--statistics lattice
calculations of the meson mass spectrum. Elastic pion--pion
scattering is discussed as an example of a complete calculation
to $O(p^6)$ in the low--energy expansion. The meson--baryon
system is the subject of the last lecture. After a short summary
of heavy baryon chiral perturbation theory, a recent analysis of
pion--nucleon scattering to $O(p^3)$ is reviewed. Finally, I
describe some very recent progress in the chiral approach to the
nucleon--nucleon interaction.
\end{abstract}
\section{The Standard Model at Low Energies}

My first Schladming Winter School took place exactly 30 years ago. 
Recalling the program of the 1968 School (Urban 1968),
many of the topics discussed at the time are still with us
today. In particular, chiral symmetry was very well represented
in 1968, with lectures by S. Glashow, F. Gursey and H. Leutwyler.
In those pre--QCD days, chiral Lagrangians were already
investigated in much detail but the prevailing understanding was
that due to their nonrenormalizability such Lagrangians could not 
be taken seriously beyond tree level. The advent of
renormalizable gauge theories at about the same time seemed to close
the chapter on chiral Lagrangians.

More than ten years later, after an influential paper of Weinberg 
(1979) and especially through the systematic analysis of Gasser and 
Leutwyler (1984, 1985), effective chiral Lagrangians were taken up 
again when it was realized that in spite of their nonrenormalizability 
they formed the basis of a consistent quantum field theory. Although 
QCD was already well established by that time
the chiral approach was shown to provide a systematic
low--energy approximation to the Standard Model in a regime
where QCD perturbation theory was obviously not applicable.

Over the years, different approaches have been pursued to investigate
the Standard Model in the low--energy domain. Most of them fall into 
the following three classes:
\begin{enumerate}
\item[i.] QCD--inspired models\\
There is a large variety of such models with more or less inspiration
from QCD. Most prominent among them are different versions of the 
Nambu--Jona-Lasinio model (Nambu and Jona-Lasinio 1961; Bijnens 1996 and
references therein) and chiral quark models (Manohar and Georgi 1984;
Bijnens et al. 1993). Those models
have provided a lot of insight into low--energy dynamics but
in the end it is difficult if not impossible to disentangle the model
dependent results from genuine QCD predictions.
\item[ii.] Lattice QCD
\item[iii.] Chiral perturbation theory (CHPT)\\
The underlying theory with quarks and gluons is replaced by an
effective field theory at the hadronic level.
Since confinement makes a perturbative matching impossible,
the traditional approach (Weinberg 1979; Gasser and Leutwyler 1984,
1985; Leutwyler 1994) relies only on the symmetries of QCD to
construct the effective field theory. The main ingredient of this
construction is the spontaneously (and explicitly) broken chiral symmetry
of QCD.
\end{enumerate} 

The purpose of these lectures is to introduce chiral symmetry as a
leit--motiv for low--energy hadron physics. The first lecture
starts with a review of spontaneous chiral symmetry breaking. In
particular, I discuss a recent classification of possible scenarios
of chiral symmetry breaking by Stern (1998) and a connection between
the quark condensate and the $V,A$ spectral functions in the
large--$N_c$ limit (Knecht and de Rafael 1997). The ingredients for
constructing the effective chiral Lagrangian of the Standard Model are
put together. This Lagrangian can be organized in two different ways
depending on the chiral counting of quark masses: standard
vs. generalized CHPT. To emphasize the importance of renormalizing
a nonrenormalizable quantum field theory like CHPT, 
the loop expansion and the renormalization procedure for the mesonic 
sector are described in some detail. After a brief review 
of quark mass ratios from CHPT, I discuss the evidence from lattice
QCD in favour of a large quark condensate. The observed
linearity of the meson masses squared as functions of the quark
masses is consistent with the standard chiral expansion to
$O(p^4)$. Moreover, it excludes small values of the quark condensate
favoured by generalized CHPT. Elastic pion--pion scattering is
considered as an example of a complete calculation to $O(p^6)$ in the 
low--energy expansion. Comparison with forthcoming experimental data
will allow for precision tests of QCD in the confinement regime. Once
again, the quark condensate enters in a crucial way. In the meson--baryon
sector, the general procedure of heavy baryon CHPT is explained for
calculating relativistic amplitudes from
frame dependent amplitudes. As an application, I review the analysis
of \cite{MM98} for elastic $\pi N$ scattering to $O(p^3)$. Finally,
some promising new developments in the chiral treatment of the
nucleon--nucleon interaction are discussed.

\subsection{Broken Chiral Symmetry}

The starting point is an idealized world where $N_f=$ 2 or 3 of the
quarks are massless ($u$, $d$ and possibly $s$). In this chiral limit,
the QCD Lagrangian
\begin{eqnarray} 
{\cal L}^0_{\rm QCD} &=& \overline{q} i \gamma^\mu\left(\partial_\mu +
i g_s {\lambda_\alpha\over 2} G^\alpha_\mu\right)q - {1\over 4}G^\alpha_
{\mu\nu}
G^{\alpha\mu\nu} + {\cal L}_{\mbox{\tiny heavy quarks}} \label{eq:QCD0}\\*
&=& \overline{q_L} i \Dslash q_L + \overline{q_R} i \Dslash q_R - 
{1\over 4}G^\alpha_
{\mu\nu}G^{\alpha\mu\nu} + {\cal L}_{\mbox{\tiny heavy quarks}}
\nonumber  \\*
q_{R,L} &=& {1\over 2}(1 \pm \gamma_5)q \hspace*{2cm}
q=\left( \begin{array}{c} u \\ d \\$[s]$ \end{array}\right) 
\nonumber 
\end{eqnarray} 
exhibits a global symmetry
$$
\underbrace{SU(N_f)_L \times SU(N_f)_R}_{\mbox{chiral group $G$}}
\times U(1)_V \times U(1)_A ~.
$$
At the effective hadronic level,  the quark
number symmetry $U(1)_V$ is realized as baryon number. The axial 
$U(1)_A$ is not a symmetry at the quantum level due to the  Abelian
anomaly ('t Hooft 1976; Callan et al. 1976; Crewther 1977) that
leads for instance to $M_{\eta'}\neq 0$ even in the chiral limit.

A classical symmetry can be realized in quantum field theory
in two different ways depending on how the vacuum responds
to a symmetry transformation. With a charge $Q=\int d^3x J^0(x)$
associated to the Noether current $J^\mu(x)$ of an internal symmetry
and for a translation invariant vacuum state $|0\rangle$, the two
realizations are distinguished by the 

\begin{center}
{\samepage
\fbox{Goldstone alternative} \\*[0.7cm]
\begin{tabular}{ccc}
$Q |0\rangle = 0$ & \hspace{3cm} & $||Q |0\rangle || = \infty$ \\*
Wigner--Weyl & & Nambu--Goldstone \\*
linear representation & & nonlinear realization \\*
degenerate multiplets & & massless Goldstone bosons \\*
exact symmetry & & spontaneously broken symmetry\\ \hline
\end{tabular}  }
\end{center}
\vspace{0.5cm}

There is compelling evidence both from phenomenology and from theory
that the chiral group G is indeed spontaneously broken~:
\begin{enumerate} 
\item[i]
Absence of parity doublets in the hadron spectrum.
\item[ii.]
The $N_f^2 -1$ pseudoscalar mesons are by far the lightest hadrons.
\item[iii.]
The vector and axial--vector spectral functions are quite
different as shown in Fig.~\ref{fig:spectral}.

\begin{figure}[t]
\centerline{\epsfig{file=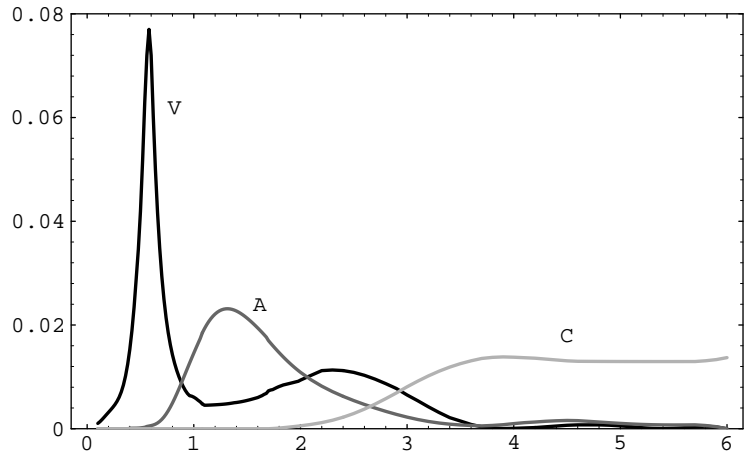,height=6cm}}
\caption{Vector and axial--vector spectral functions in the $I=1$
channel as functions of
s (in GeV$^2$) from Donoghue and Perez (1997).
$V,A$ stand for the isovector resonance contributions and $C$ denotes 
the (common) continuum contribution.}
\label{fig:spectral}
\end{figure}

\item[iv.]
The anomaly matching conditions ('t Hooft 1980; Frishman et al.
1981; Coleman and Grossman 1982) together with confinement require the
spontaneous breaking of G for $N_f\ge 3$. 
\item[v.]
In vector--like gauge theories like QCD (with the vacuum angle 
$\theta_{QCD}=0$), vector symmetries like the diagonal subgroup
of $G$, $SU(N_f)_V$, remain unbroken (Vafa and Witten 1984).
\item[vi.]
There is by now overwhelming evidence from lattice gauge theories (see
below) for a nonvanishing quark condensate.
\end{enumerate} 

All these arguments together suggest very strongly that the chiral 
symmetry $G$ is spontaneously broken to the vectorial subgroup 
$SU(N_f)_V$ (isospin for $N_f=2$, flavour $SU(3)$ for $N_f=3$):
\begin{equation} 
G \longrightarrow H=SU(N_f)_V~.
\end{equation} 

To investigate the underlying mechanism further,
let me recall one of the standard proofs of the Goldstone theorem
(Goldstone 1961): starting with the charge operator in a finite volume $V$, 
$Q^V=\int_V d^3x J^0(x)$, one assumes the existence of a (local)
operator $A$ such that
\begin{equation} 
\lim_{V\rightarrow  \infty} \langle 0|[Q^V(x^0),A]|0\rangle \neq 0 ~,
\label{eq:op}
\end{equation} 
which is of course only possible if
\begin{equation} 
  Q |0\rangle \neq 0 ~.
\end{equation} 
Then the Goldstone theorem tells us that there exists a massless
state $|G\rangle$ with 
\begin{equation} 
\langle 0|J^0(0)|G\rangle\langle G|A|0\rangle \neq 0 ~.
\label{eq:Gth}
\end{equation} 

The left--hand side of Eq.~(\ref{eq:op}) is called an order parameter
of the spontaneous symmetry breaking. The relation (\ref{eq:Gth})
contains two nonvanishing matrix elements. The first one 
involves only the symmetry current and it is therefore
independent of the specific order parameter:
\begin{equation}
\langle 0|J^0(0)|G\rangle \neq 0 \label{eq:Gme}
\end{equation}
is a necessary and sufficient condition for spontaneous
breaking. The second matrix element in (\ref{eq:Gth}), on the other 
hand, does depend on
the order parameter considered. Together with (\ref{eq:Gme}), its 
nonvanishing is sufficient but
of course not necessary for the Nambu--Goldstone mechanism.

In QCD, the charges in question are the axial charges
\begin{equation} 
Q^i_A = Q^i_R-Q^i_L  \qquad (i=1,\dots,N_f^2-1)~.
\end{equation} 
Which is (are) the order parameter(s) of spontaneous chiral symmetry
breaking in QCD? From the discussion above, we infer that the
operator $A$ in (\ref{eq:op}) must be a colour--singlet,
pseudoscalar quark--gluon operator. The unique choice for a local
operator in QCD with lowest operator dimension three
is\footnote{Here, the $\lambda_i$ are the generators of $SU(N_f)_V$
in the fundamental representation.}
\begin{equation} 
A_i = \overline{q}\gamma_5 \lambda_i q
\end{equation} 
with
\begin{equation} 
\left[Q^i_A,A_j\right]=-\displaystyle\frac{1}{2} \overline{q}
\{\lambda_i,\lambda_j\} q ~.
\end{equation} 
If the vacuum is invariant under $SU(N_f)_V$, 
\begin{equation} 
\langle 0|\overline{u}u|0\rangle=\langle 0|\overline{d}d|0\rangle
\left[=\langle 0|\overline{s}s|0\rangle\right]~.
\end{equation} 
Thus, a nonvanishing quark condensate
\begin{equation} 
\langle 0|\overline{q}q|0\rangle \neq 0 \label{eq:qcond}
\end{equation} 
is sufficient for spontaneous chiral symmetry
breaking. As already emphasized, (\ref{eq:qcond})
is certainly not a necessary condition. Increasing the
operator dimension, the next candidate is the so--called mixed
condensate of dimension five,
\begin{equation} 
\langle 0|\overline{q}\sigma_{\mu\nu}\lambda_\alpha q G^{\alpha\mu\nu}
|0\rangle \neq 0~,
\end{equation} 
and there are many more possibilities for operator dimensions 
$\ge 6$. All order parameters are in principle equally good for
triggering the Goldstone mechanism. As we will see later on, the
quark condensate enjoys nevertheless a special status. Although the
following statement will have to be made more precise, we are going to
investigate whether the quark condensate is the dominant order 
parameter of spontaneous chiral symmetry breaking in QCD.

To analyse the possible scenarios, it is useful to consider QCD in a
Euclidean box of finite volume $V=L^4$. The Lagrangian for a massive
quark in a given gluonic background is 
\begin{equation} 
\cL = \overline{q}(\Dslash + m) q
\end{equation} 
with hermitian $i\Dslash$. In a finite volume, the Dirac operator has a
discrete spectrum~:
\begin{equation} 
i\Dslash u_n = \lambda_n u_n
\end{equation} 
with real eigenvalues $\lambda_n$ and orthonormal spinorial
eigenfunctions $u_n$. Spontaneous chiral symmetry breaking is
related to the infrared structure of this spectrum in the limit 
$V\to \infty$ (Banks and Casher 1980; Vafa and Witten 1984; Leutwyler
and Smilga 1992; \dots; Stern 1998).

The main reason for working in Euclidean space is the following. 
Because of
\begin{equation} 
i\Dslash u_n = \lambda_n u_n \longrightarrow i\Dslash \gamma_5 u_n = 
- \lambda_n \gamma_5 u_n ~,
\end{equation} 
the nonzero eigenvalues come in pairs $\pm \lambda_n$. 
Therefore, the fermion determinant
in a given gluon background is real and positive (for
$\theta_{QCD}=0$)~:
\begin{equation} 
\det(\Dslash + m)= m^\nu \displaystyle\prod_{\lambda_n\neq 0}(m-
i\lambda_n)=
m^\nu \displaystyle\prod_{\lambda_n > 0}(m^2+\lambda_n^2) > 0~,
\end{equation}
where $\nu$ is the multiplicity of the zero modes. The
fermion integration yields a real, positive measure for the gluonic
functional integral. Thus, many statements for correlation functions
in a given gluon background will survive the functional average
over the gluon fields.

The quark two--point function for coinciding arguments can be written
as (the subscript $G$ denotes the gluon background)
\begin{equation} 
\langle \overline{q}(x) q(x)\rangle_G = - \displaystyle\sum_n
\displaystyle\frac{u_n^\dagger(x)u_n(x)}{m-i\lambda_n}
\end{equation} 
implying\footnote{The zero modes will not be relevant in the 
infinite--volume limit.}
\begin{equation} 
\displaystyle\frac{1}{V}\displaystyle\int d^4x \langle \overline{q}(x)
q(x)\rangle_G = -\displaystyle\frac{1}{V} \displaystyle\sum_n
\displaystyle\frac{1}{m-i\lambda_n}=
 -\displaystyle\frac{2m}{V} \displaystyle\sum_{\lambda_n>0}
\displaystyle\frac{1}{m^2+\lambda_n^2}~.\label{eq:qqV}
\end{equation} 
This relation demonstrates that the chiral and
the infinite--volume limits do not commute. Taking the chiral limit
$m\to 0$ for fixed volume yields $\langle \overline{q}q\rangle_G=0$,
in accordance with the fact that there is no spontaneous
symmetry breaking in a finite volume. The limit of interest is
therefore first $V\to \infty$ for fixed $m$ and then $m\to 0$.

For $V\to \infty$, the eigenvalues $\lambda_n$ become dense and we
must replace the sum over eigenvalues by an integral over a density
$\rho(\lambda)$:
$$
\displaystyle\frac{1}{V}\sum_n \stackrel{{\tiny V\to \infty}}
{\longrightarrow} \int d\lambda \rho(\lambda)~.
$$
Averaging the relation (\ref{eq:qqV}) over gluon fields and
taking the infinite--volume limit, one gets
\begin{equation}
\langle 0|\overline{q}q|0\rangle = - \displaystyle\int_{-\infty}
^{\infty}\displaystyle\frac{d\lambda \rho(\lambda)}
{m-i\lambda}= - 2m\displaystyle\int_{0}
^{\infty}\displaystyle\frac{d\lambda \rho(\lambda)}
{m^2+\lambda^2}~.
\end{equation}
In the chiral limit, we obtain the relation of Banks and Casher
(1980)~: 
\begin{equation}
\lim_{m\to 0} \langle 0|\overline{q}q|0\rangle = -\pi \rho(0)~.
\end{equation}
For free fields, $\rho(\lambda)\sim \lambda^3$ near 
$\lambda=0$. Thus, the eigenvalues must accumulate near zero
to produce a nonvanishing quark condensate. Although the Banks--Casher
relation does not tell us which gauge field configurations could be
responsible for $\rho(0)\neq 0$, many suggestions are on the market
(instantons, monopoles, \dots).

This is a good place to recall the gist of
the Vafa--Witten argument for the conservation of vector symmetries
(Vafa and Witten 1984):
\begin{eqnarray}
\langle 0|\overline{u}u-\overline{d}d|0\rangle &=&
-\displaystyle\int_{-\infty}^{\infty}d\lambda 
\rho(\lambda)\left(\displaystyle\frac{1}{m_u-i\lambda}-
\displaystyle\frac{1}{m_d-i\lambda}\right)\nonumber\\
&=&(m_u-m_d)\displaystyle\int_{-\infty}^{\infty}
\displaystyle\frac{d\lambda \rho(\lambda)}{(m_u-i\lambda)
(m_d-i\lambda)}\nonumber\\
&\stackrel{m_u\to m_d}{\longrightarrow}& 0~.\label{eq:VW}
\end{eqnarray}
Unlike in the chiral limit, the integrand in (\ref{eq:VW}) does
not become singular in the equal--mass limit and the vacuum
remains $SU(N_f)_V$ invariant.

The previous discussion concentrated on one specific order
parameter for spontaneous chiral symmetry breaking, the quark
condensate. Stern (1998) has recently performed a similar
analysis for a quantity that is directly related to the
Goldstone matrix element (\ref{eq:Gme}). Consider the
correlation function
\begin{equation}
\Pi^{\mu\nu}_{LR}(q) \delta_{ij}=4 i\displaystyle\int d^4x e^{iqx}
\langle 0|T L^\mu_i(x) R^\nu_j(0)|0\rangle \label{eq:LR}
\end{equation}
$$
L^\mu_i=\overline{q_L}\gamma^\mu\frac{\lambda_i}{2} q_L ~, \qquad
R^\mu_i=\overline{q_R}\gamma^\mu\frac{\lambda_i}{2} q_R~.
$$
In the chiral limit, the correlator vanishes for any $q$ unless
the vacuum is asymmetric. In particular, one finds in the chiral limit
\begin{equation}
\lim_{m_q\to 0} \Pi^{\mu\nu}_{LR}(0)=-F^2 g^{\mu\nu}
\end{equation}
where the constant $F$ (the pion decay constant in the
chiral limit) characterizes the Goldstone matrix
element (\ref{eq:Gme}):
\begin{equation}
\langle 0|\overline{q}\gamma^\mu\gamma_5\frac{\lambda_i}{2} q|\varphi_j(p)
\rangle=i\delta_{ij} F \left[1+O(m_q)\right]p^\mu e^{-ipx}~.
\end{equation}
Thus, $\Pi^{\mu\nu}_{LR}(0)\neq 0$ is a necessary and
sufficient condition for spontaneous chiral symmetry breaking.

Introducing the average (over all gluon configurations) number
of states $N(\varepsilon,L)$  with $|\lambda|\le\varepsilon$,
Stern (1998) defines a mean eigenvalue density $\widehat\rho$ in finite
volume as
\begin{equation} 
\widehat\rho(\varepsilon,L) =
\displaystyle\frac{N(\varepsilon,L)}{2\varepsilon V}~.
\end{equation} 
Of course,
\begin{equation} 
\rho(0)=\lim_{\varepsilon\to 0}\lim_{L\to
\infty}\widehat\rho(\varepsilon,L) 
\end{equation} 
with the previously introduced density $\rho$.
With similar techniques as before (again in Euclidean space),
Stern (1998) has derived a relation for the decay constant F~:
\begin{equation}
F^2=\pi^2\lim_{\varepsilon\to 0}\lim_{L\to \infty}L^4 J(\varepsilon,L) 
\widehat\rho(\varepsilon,L)^2
\label{eq:Kubo}
\end{equation}
in terms of an average transition probability between states with
$|\lambda|\le\varepsilon$:
\begin{equation} 
J(\varepsilon,L)=\displaystyle\frac{1}{N(\varepsilon,L)^2}
<<\displaystyle\sum_{kl}^\varepsilon J_{kl}>>~,\quad
J_{kl}=\displaystyle\frac{1}{4}\displaystyle\sum_\mu|\int d^4x 
u_k^\dagger(x)\gamma_\mu u_l(x)|^2
\end{equation} 
where $<<\dots>>$ denotes an average over gluon configurations. The
formula (\ref{eq:Kubo}) closely resembles the Greenwood--Kubo formula
for electric conductivity (see Stern 1998). 

As already emphasized, the eigenvalues $\lambda_n$ must
accumulate near zero to trigger spontaneous chiral symmetry
breaking. A crucial parameter is the critical exponent
$\kappa$ defined as (Stern 1998)
\begin{equation} 
<<\lambda_n>> \sim L^{-\kappa} \label{eq:crit}
\end{equation} 
for $\lambda_n$ near zero and $L\to \infty$. Up to higher powers 
in $\varepsilon$, the
average number of states and the mean eigenvalue density depend on 
$\kappa$ as
\begin{eqnarray}
N(\varepsilon,L) &=& \left(\displaystyle\frac{2\varepsilon}{\mu}\right)
^{\displaystyle\frac{4}{\kappa}}(\mu L)^4 +\dots \label{eq:Nrho1}\\
\widehat\rho(\varepsilon,L) &=& \left(\displaystyle\frac{2\varepsilon}{\mu}\right)
^{\displaystyle\frac{4}{\kappa}-1}\mu^3 +\dots \label{eq:Nrho2}
\end{eqnarray}
in terms of some energy scale $\mu$. As is obvious from the definition
(\ref{eq:crit}) and from the expressions
(\ref{eq:Nrho1}),(\ref{eq:Nrho2}), the
eigenvalues with maximal $\kappa$ are the relevant ones. 
The completeness sum rule $\sum_l J_{kl}=1$ for the transition
probabilities yields an upper bound for $F^2$ (Stern 1998)~:
\begin{equation}
F^2 \le\pi^2 \mu^2 \lim_{\varepsilon\to 0}\left(\displaystyle\frac
{2\varepsilon}{\mu}\right)^{\displaystyle\frac{4}{\kappa}-2}~.
\label{eq:csr}
\end{equation}
Therefore, while $\kappa=1$ for free fields, spontaneous chiral
symmetry breaking requires $\kappa \ge 2$. With the same
notation, we also have
\begin{equation}
\langle 0|\overline{q}q|0\rangle =-\pi \mu^3 
\lim_{\varepsilon\to 0}\left(\displaystyle\frac{2\varepsilon}{\mu}\right)
^{\displaystyle\frac{4}{\kappa}-1}
\end{equation}
leading to $\kappa=4$ for a nonvanishing quark condensate (Leutwyler
and Smilga 1992). On rather general grounds, the critical index is
bounded by
\begin{equation} 
1 \le \kappa \le 4~.
\end{equation} 
Stern (1998) has argued that the existence of an effective
chiral Lagrangian analytic in the quark masses suggests that the
exponent $4/\kappa$ is actually an integer\footnote{There are 
explicit counterexamples to this analyticity
assumption in less than four dimensions (L. Alvarez-Gaum\'e, H.
Grosse and J. Stern, private communications).}. In this case, only
$\kappa=1$ or $\kappa=2, 4$ would be allowed,
the latter two cases being compatible with spontaneous chiral
symmetry breaking.

There are then two preferred scenarios for spontaneous chiral symmetry
breaking (Stern 1998):
\begin{enumerate}
\item[i.] $\kappa=2~:$ \\
The density of states near $\varepsilon=0$ is too small to generate a
nonvanishing quark condensate, but the high ``quark mobility'' $J$
induces $F\neq 0$.
\item[ii.] $\kappa=4~:$ \\
Here, the density of states is
sufficiently large for $\rho(0)\neq 0$. This option
is strongly supported by lattice data (see below) favouring a
nonvanishing quark condensate. With hindsight, the scenario
most likely realized in nature is at least consistent with the 
previous analyticity hypothesis.
\end{enumerate}

Are there other indications for a large quark condensate?
Knecht and de Rafael (1997) have
recently found an interesting relation between chiral order
parameters and the vector and axial--vector spectral functions in the
limit of large $N_c$. They consider again the correlation function 
(\ref{eq:LR}). In the chiral limit, it can be expressed in terms of a
single scalar function $\Pi_{LR}(Q^2)$~:
\begin{equation} 
\Pi^{\mu\nu}_{LR}(q)=(q^\mu q^\nu-g^{\mu\nu}q^2)\Pi_{LR}(Q^2)~,\qquad
Q^2=-q^2~.
\end{equation} 
Because it is a (nonlocal) order parameter, $\Pi_{LR}(Q^2)$
vanishes in all orders of QCD perturbation theory for a
symmetric vacuum. The asymptotic behaviour for large and small
$Q^2$ ($Q^2 \ge 0$) is 
\begin{eqnarray}
\Pi_{LR}(Q^2)&=&-\displaystyle\frac{4\pi}{Q^6}[\alpha_s + O(\alpha_s^2)]
\langle \overline{u}u\rangle^2+O(\displaystyle\frac{1}{Q^8}) 
\label{eq:SVZ}\\
-Q^2 \Pi_{LR}(Q^2)&=& F^2 + O(Q^2)~.
\end{eqnarray}
For the large--$Q^2$ behaviour (\ref{eq:SVZ}) (Shifman et al.
1979), $N_c\to\infty$ has already been assumed to factorize the
four--quark condensate into the square of the (two--)quark condensate.
In the same limit, the correlation function $\Pi_{LR}(Q^2)$
is determined by an infinite number of stable vector and
axial--vector states:
\begin{equation}
-Q^2 \Pi_{LR}(Q^2) = F^2 + \sum_A\displaystyle\frac{F_A^2 Q^2}{M_A^2+Q^2}-
\sum_V\displaystyle\frac{F_V^2 Q^2}{M_V^2+Q^2}~,\label{eq:VAun}
\end{equation}
where $M_I, F_I (I=V,A)$ are the masses and the coupling
strengths of the spin--1 mesons to the respective currents.
Comparison with the asymptotic behaviour (\ref{eq:SVZ}) yields
the two Weinberg sum rules (Weinberg 1967)
\begin{eqnarray}
\sum_V F_V^2 - \sum_A F_A^2=F^2 \label{eq:WSR1} \\
\sum_V F_V^2 M_V^2 - \sum_A F_A^2 M_A^2 = 0 \label{eq:WSR2}
\end{eqnarray}
and allows (\ref{eq:VAun}) to be rewritten as
\begin{equation}
-Q^2 \Pi_{LR}(Q^2) = \sum_A\displaystyle\frac{F_A^2 M_A^4}{Q^2(M_A^2+Q^2)}-
\sum_V\displaystyle\frac{F_V^2 M_V^4}{Q^2(M_V^2+Q^2)}~.\label{eq:VAsub}
\end{equation}
This expression can now be matched once more to the asymptotic
behaviour (\ref{eq:SVZ}). Referring to Knecht and de Rafael
(1997) for a general discussion, I concentrate here on the
simplest possibility assuming that the $V,A$ spectral functions can be
described by single resonance states plus a continuum. The
experimental situation for the $I=1$ channel shown in
Fig.~\ref{fig:spectral} is
clearly not very far from this simplest case. In addition to the
inequality $M_V < M_A$ following from the Weinberg sum rules
(\ref{eq:WSR1}), (\ref{eq:WSR2}), the matching condition requires
\begin{equation} 
4\pi [\alpha_s + O(\alpha_s^2)] \langle \overline{u}u\rangle^2=
F^2 M_V^2 M_A^2
\end{equation} 
or approximately
\begin{equation}
4\pi \alpha_s \langle \overline{u}u\rangle^2 \simeq
F_\pi^2 M_\rho^2 M_{A_1}^2 ~.
\end{equation}
From the last relation, Knecht and de Rafael (1997) extract a
quark condensate 
\begin{equation} 
\langle \overline{u}u\rangle(\nu=1~{\rm GeV}) \simeq - (303 {\rm ~MeV})^3
\end{equation} 
with $\nu$ the QCD renormalization scale in the
$\overline{\rm MS}$ scheme.
In view of the assumptions made, especially the large--$N_c$
limit, this value is quite compatible with
\begin{equation}
\langle \overline{u}u\rangle(\nu=1~{\rm GeV}) = -\left[
(229 \pm 9) {\rm ~MeV}\right]^3 \label{eq:DN}
\end{equation}
from a recent compilation of sum rule estimates (Dosch and Narison 
1998).

The conclusion is that the $V,A$ spectrum is fully consistent
with both sum rule and lattice estimates for the quark condensate.
We come back to this issue in the discussion of light quark masses.

\subsection{Effective Field Theory}

The pseudoscalar mesons are not only the lightest hadrons but
they also have a special status as (pseudo--) Goldstone bosons.
In the chiral limit, the interactions of Goldstone bosons vanish
as their energies tend to zero. In other words, the interactions
of Goldstone bosons become arbitrarily weak for decreasing
energy no matter how strong the underlying interaction is. This
is the basis for a systematic low--energy expansion with an effective
chiral Lagrangian that is organized in a derivative expansion.

There is a standard procedure for implementing a symmetry
transformation on Goldstone fields (Coleman et al. 1969;
Callan et al. 1969). Geometrically, the Goldstone fields $\varphi=\pi
[,K,\eta_8]$ can be viewed as coordinates of the coset space $G/H$. 
They are assembled in a matrix field $u(\varphi)\in G/H$, the basic
building block of chiral Lagrangians. Different forms of this
matrix field (e.g., the exponential representation) correspond to 
different parametrizations of coset space. Since the chiral 
Lagrangian is generically nonrenormalizable, there is no 
distinguished choice of field variables as for 
renormalizable quantum field theories.

An element $g$ of the symmetry group $G$ 
induces in a natural way a transformation of $u(\varphi)$ by left
translation:
\begin{equation} 
u(\varphi) \stackrel{g\in G}{\longrightarrow} g u(\varphi) = 
u(\varphi') h(g,\varphi)~.  \label{eq:coset}
\end{equation}
The so--called compensator field $h(g,\varphi)$ is an element of the conserved
subgroup $H$ and it accounts for the fact that a coset element is
only defined up to an $H$ transformation. For $g \in H$, the
symmetry is realized in the usual linear way (Wigner--Weyl) and $h(g)$
does not depend on the Goldstone fields $\varphi$. On the other hand,
for $g \in G$ corresponding to a spontaneously broken symmetry 
($g \not\in H$), the symmetry is realized nonlinearly (Nambu--Goldstone)
and $h(g,\varphi)$ does depend on $\varphi$.

For the special case of chiral symmetry $G=SU(N_f)_L \times SU(N_f)_R$,
parity relates left-- and right--chiral transformations. With a standard
choice of coset representatives, the general transformation 
(\ref{eq:coset}) takes the special form
\begin{equation}
 u(\varphi')=g_R u(\varphi) h(g,\varphi)^{-1} 
= h(g,\varphi) u(\varphi) g_L^{-1} \label{eq:uphi}
\end{equation}
$$
g=(g_L,g_R) \in G~.
$$

For practical purposes, one never needs to know the explicit form
of $h(g,\varphi)$, but only the transformation property
(\ref{eq:uphi}). In the mesonic sector, it is often more convenient
to work with the square of $u(\varphi)$. Because of (\ref{eq:uphi}), the 
matrix field $U(\varphi)=u(\varphi)^2$ has a simpler linear 
transformation behaviour:
\begin{equation}
U(\varphi) \stackrel{G}{\to} g_R U(\varphi) g_L^{-1} \label{eq:Uphi}~.
\end{equation}
It is therefore frequently used as basic building block for 
mesonic chiral Lagrangians. 

When non--Goldstone degrees of freedom like baryons or
meson resonances are included in the effective Lagrangians, the 
nonlinear picture with $u(\varphi)$ and $h(g,\varphi)$ is more 
appropriate. If a generic hadron field $\Psi$ (with $M_\Psi \neq 0$ 
in the chiral limit) transforms under $H$ as
\begin{equation} 
\Psi \stackrel{h\in H}{\to} \Psi'=h_\Psi (h) \Psi  
\end{equation} 
according to a given representation $h_\Psi$ of $H$, the
compensator field in this representation furnishes immediately
a realization of all of G:
\begin{equation}
\Psi \stackrel{g\in G}{\to} \Psi'=h_\Psi(g,\varphi) \Psi  ~.
\end{equation}
This transformation is not only nonlinear in $\varphi$ but also
space--time dependent requiring the introduction of a chirally
covariant derivative. We will come back to this case in the last
lecture on baryons and mesons. 

Before embarking on the construction of an effective field
theory for QCD, we pause for a moment to realize that there is
in fact no chiral symmetry in nature. In addition to the
spontaneous breaking discussed so far, chiral symmetry is broken
explicitly both by nonvanishing quark masses and by the electroweak
interactions of hadrons. 

The main assumption of CHPT is that it
makes sense to expand around the chiral limit. In full generality,
chiral Lagrangians are therefore constructed by means of a
two--fold expansion in both
\begin{itemize}
\item derivatives ($\sim$ momenta) and
\item quark masses~:
\end{itemize}
\begin{equation}
\cL_{\rm eff} = \sum_{i,j} \cL_{ij}~, \qquad \qquad
\cL_{ij} = O(p^i m^j_q)~. \label{eq:gexp}
\end{equation}
The two expansions become related by expressing the pseudoscalar meson
masses in terms of the quark masses $m_q$. If the quark condensate is
nonvanishing in the chiral limit, the squares of the meson masses
start out linear in $m_q$ (see below). The constant of proportionality 
is a quantity $B$ with 
\begin{equation} 
B=-\displaystyle\frac{\langle\overline{u}u\rangle}{F^2}
\end{equation} 
in the chiral limit. Assuming the linear terms to provide the dominant 
contributions to the meson masses corresponds to a scale (the
product $B m_q$ is scale invariant)
\begin{equation}
B(\nu = 1 {\rm ~GeV})\simeq 1.4 {\rm ~GeV} ~.
\end{equation}
This standard scenario of CHPT (Weinberg 1979; Gasser and
Leut\-wy\-ler 1984, 1985; Leut\-wy\-ler 1994) is compatible with a 
large quark condensate as given for instance in (\ref{eq:DN}). 
The standard chiral counting
\begin{equation} 
m_q=O(M^2)=O(p^2)
\end{equation} 
reduces the two--fold expansion (\ref{eq:gexp}) to
\begin{equation}
\cL_{\rm eff} = \sum_n \cL_n~, \qquad \qquad
\cL_n = \sum_{i + 2j = n} \cL_{ij}~.\label{eq:Leff}
\end{equation}
For mesons, the chiral expansion proceeds in steps of two ($n=$
2,4,6,\dots) because the index $i$ is even.

Despite the evidence in favour of the standard scenario, the 
alternative of a much smaller or even vanishing quark condensate
(e.g., for $\kappa=2$ in the previous classification of chiral
symmetry breaking) is actively
being pursued (Fuchs et al. 1991; Stern et al. 1993; Knecht et
al. 1993, 1995; Stern 1997 and references therein). This
option is characterized by
\begin{equation}
B(\nu = 1 {\rm ~GeV}) \sim O(F_\pi) 
\end{equation}
with the pion decay constant $F_\pi=92.4$ MeV. The so--called
generalized CHPT amounts to a reordering 
of the effective chiral Lagrangian (\ref{eq:Leff}) on the basis of a 
modified chiral counting with $m_q=O(p)$. We will come back to
generalized CHPT in
several instances, in particular during the discussion of quark masses,
but for most of these lectures I will stay with the mainstream
of standard CHPT.

Both conceptually and for practical purposes, the best way to keep track
of the explicit breaking is through the introduction of external matrix
fields (Gasser and Leutwyler 1984, 1985) $v_\mu,a_\mu,s,p$. The QCD
Lagrangian (\ref{eq:QCD0}) with $N_f$ massless quarks is extended to
\begin{equation} 
\cL = \cL^0_{\rm QCD} + \overline q \gamma^\mu(v_\mu + a_\mu \gamma_5)q -
\overline q (s - ip \gamma_5)q \label{eq:QCD}
\end{equation} 
to include electroweak interactions of quarks with external gauge
fields $v_\mu,a_\mu$ and to allow for nonzero quark masses by setting 
the scalar matrix field $s(x)$ equal to the diagonal quark mass
matrix. The big advantage is that one can perform all calculations
with a (locally) $G$ invariant effective
Lagrangian in a manifestly chiral invariant manner. Only at the very 
end, one inserts the appropriate
external fields to extract the Green functions of quark
currents or matrix elements of interest. The explicit breaking of
chiral symmetry is automatically taken care of by this spurion technique.
In addition, electromagnetic gauge invariance is manifest.

Although this procedure produces all Green functions for electromagnetic
and weak currents, the method must be extended in order to include
virtual photons (electromagnetic corrections) or virtual $W$ bosons
(nonleptonic weak interactions). The present status of the effective
chiral Lagrangian of the Standard Model is summarized in Table
\ref{tab:ecL}. The purely mesonic Lagrangian is denoted as
$\cL_2$+$\cL_4$+$\cL_6$ and will be discussed at length in the
following lecture. Even (odd) refers to terms in the meson
Lagrangian without (with) an
$\varepsilon$ tensor. The pion--nucleon Lagrangian $\sum_n \cL_n^{\pi N}$
will be the subject of the last lecture. The chiral Lagrangians for
virtual photons (superscript $\gamma$) and for nonleptonic weak
interactions (superscript $\Delta S=1$) will not be treated in
these lectures. The
numbers in brackets denote the number of independent coupling
constants or low--energy constants (LECs) for the given
Lagrangian. They apply in general for $N_f=3$ except for the $\pi N$
Lagrangian ($N_f=2$) and for the mesonic Lagrangian of $O(p^6)$
(general $N_f$). The different Lagrangians are grouped together
according to the chiral order that corresponds to the indicated loop
order. The underlined parts denote completely renormalized Lagrangians.

\renewcommand{\arraystretch}{1.1}
\begin{table}[t]
\begin{center}
\caption{The effective chiral Lagrangian of the Standard Model}
\label{tab:ecL}
\vspace{.5cm}
\begin{tabular}{lc|c} 
\hspace{2cm} ${\cal L}_{\rm chiral \,\, dimension}$ ~($\#$ of LECs)  
& \hspace{.2cm} &\hspace{.1cm} loop order \\[10pt] 
\hline 
& & \\[10pt]
${\cal L}_2(2)$~+~${\cal L}_4^{\rm odd}(0)$~+~${\cal L}_2^{\Delta S=1}(2)$
~+~${\cal L}_0^\gamma(1)$  & \hspace{.2cm} & $L=0$ \\[10pt]
~+~${\cal L}_1^{\pi N}(1)$~+~${\cal L}_2^{\pi N}(7)$~+~\dots & & \\[20pt]
~+~$\underline{{\cal L}_4^{\rm even}(10)}$~+~$\underline{{\cal
L}_6^{\rm odd}(32)}$
~+~$\underline{{\cal L}_4^{\Delta S=1}(22,{\rm octet})}$~+
~$\underline{{\cal L}_2^\gamma(14)}$ & \hspace{.3cm} &  $L=1$ \\[10pt]
~+~$\underline{{\cal L}_3^{\pi N}(23)}$~+~${\cal L}_4^{\pi
N}(?)$~+~\dots & & \\[20pt]
~+~$\underline{{\cal L}_6^{\rm even}(112~{\rm for}~SU(N_f))}$~+
~\dots & & $L=2$ \\[12pt] \hline
\end{tabular}
\end{center}
\end{table}

A striking feature of Table \ref{tab:ecL} is the rapidly
growing number of LECs with increasing chiral order. Those
constants describe the influence of all states that are not
represented by 
explicit fields in the effective chiral Lagrangians. Although the
general strategy of CHPT has been to fix those constants from experiment 
and then make predictions for other observables there is obviously a
natural limit for such a program. This is the inescapable consequence
of a nonrenormalizable effective Lagrangian that is constructed solely
on the basis of symmetry considerations. Nevertheless, I will try to
convince you that even with 112 coupling constants  
one can make reliable predictions for low--energy observables.

\section{Chiral Perturbation Theory with Mesons}

The effective chiral Lagrangian for the strong
interactions of mesons is constructed in terms of the basic building
blocks $U(\varphi)$ and the external fields $v_\mu$, $a_\mu$, 
$s$ and $p$. With the standard chiral counting
described previously, the chiral Lagrangian starts
at $O(p^2)$ with
\begin{equation} 
\cL_2 = \displaystyle\frac{F^2}{4} \langle D_\mu U D^\mu U^\dagger + 
\chi U^\dagger + \chi^\dagger  U \rangle \label{eq:L2}
\end{equation} 
$$ 
\chi = 2B(s + ip) \qquad \qquad 
D_\mu U = \partial_\mu U - i  (v_\mu + a_\mu) U + 
i U (v_\mu - a_\mu)
$$
where $\langle \dots \rangle$ stands for the $N_f-$dimensional trace.
We have already encountered both LECs of $O(p^2)$. They are related to
the pion decay constant and to the quark condensate:
\begin{equation} 
F_\pi =  F[1 + O(m_q)] = 92.4 ~{\rm MeV} \label{eq:FB}
\end{equation} 
$$\langle 0|\bar u u |0\rangle = - F^2 B[1 + O(m_q)] ~.
$$
Expanding the Lagrangian (\ref{eq:L2}) to second order in the meson
fields and setting the external scalar field equal to the quark mass 
matrix, one can immediately read off the pseudoscalar meson masses
to leading order in $m_q$, e.g.,
\begin{equation} 
M^2_{\pi^+}  = (m_u + m_d) B~.\label{eq:mpi2}
\end{equation} 
As expected, for $B \neq 0$ the squares of the meson masses are linear 
in the quark masses to leading order. The full set of
equations ($N_f=3$) for the masses
of the pseudoscalar octet gives rise to several well--known relations:
\begin{eqnarray} 
F_\pi^2 M_\pi^2 = - (m_u + m_d) \langle 0|\bar u u|0\rangle &
\mbox{      } & \mbox{ (Gell-Mann et al. 1968) } \label{eq:GMOR} \\
\displaystyle\frac{M_\pi^2}{m_u+m_d} = \displaystyle\frac
{M_{K^+}^2}{m_s + m_u} =
\displaystyle\frac{M_{K^0}^2}{m_s + m_d} & \mbox{      } & 
\mbox{ (Weinberg 1977) } \label{eq:ratios} \\
3M^2_{\eta_8} = 4 M_K^2 - M_\pi^2 & \mbox{      } & 
\mbox{ (Gell-Mann 1957; Okubo 1962) } 
\label{eq:GMO}
\end{eqnarray} 

Having determined the two LECs of $O(p^2)$, we may now calculate from
the Lagrangian (\ref{eq:L2}) any Green function or S--matrix amplitude
without free parameters. The resulting tree--level amplitudes are the
leading expressions in the low--energy expansion of the Standard
Model. They are given in terms of $F_\pi$ and meson masses and
they correspond to the current algebra amplitudes of the sixties if we
adopt the standard chiral counting. 

The situation becomes more involved once we go to next--to--leading
order, $O(p^4)$. Before presenting the general procedure, we observe that
no matter how many higher--order Lagrangians we include, tree amplitudes
will always be real. On the other hand, unitarity and analyticity
require complex amplitudes in general. A good example is elastic
pion--pion scattering where the partial--wave amplitudes $t^I_l(s)$
satisfy the unitarity constraint
\begin{equation} 
\Im m ~t^I_l(s) \ge (1-\displaystyle\frac{4 M_\pi^2}{s})^{\frac{1}{2}} 
|t^I_l(s)|^2~.
\end{equation} 
Since $t^I_l(s)$ starts out at $O(p^2)$ (for $l<2$), the
partial--wave amplitudes are complex from $O(p^4)$ on. 

This example illustrates the general requirement that a systematic
low--energy expansion entails a loop expansion. Since loop amplitudes
are in general divergent, regularization and renormalization are
essential ingredients of CHPT. Any regularization is in principle
equally acceptable, but dimensional regularization is the most popular
method for well--known reasons.

Although the need for regularization is beyond debate, the situation
is more subtle concerning renormalization. Here are two recurrent 
questions in this connection:
\begin{itemize} 
\item Why bother renormalizing a quantum field theory that is
after all based on a nonrenormalizable Lagrangian?
\item Why not use a ``physical'' cutoff instead?
\end{itemize} 
The answer to both questions is that we are interested in predictions 
of the Standard Model itself rather than of some cutoff version
no matter how ``physical'' that cutoff may be. Renormalization
ensures that the final results are independent of the chosen
regularization method. As we will now discuss in some detail,
renormalization amounts to absorbing the divergences in the LECs of
higher--order chiral Lagrangians. The renormalized LECs are then
measurable, although in general scale dependent quantities. In any
physical amplitude, this scale dependence always cancels the
scale dependence of loop amplitudes.

\subsection{Loop Expansion and Renormalization}

This part of the lectures is on a more technical level than the
rest. Its purpose is to demonstrate that we are taking
the quantum field theory aspects of chiral Lagrangians seriously.

The strong interactions of mesons are described
by the generating functional of Green functions (of quark currents)
\begin{equation} 
e^{\displaystyle i Z[j]} = <0~{\rm out}|0~{\rm in}>_{j}~ = 
\int [d\varphi] e^{\displaystyle i S_{\rm eff}[\varphi,j]}
\end{equation} 
where
$$ j \sim v,a,s,p $$
denotes collectively the external fields.

The chiral expansion of the action
\begin{eqnarray}  
S_{\rm eff}[\varphi,j] &=& S_2[\varphi,j] + S_4[\varphi,j] + 
S_6[\varphi,j] + \dots \\*
S_n[\varphi,j]&=&\int d^4x \cL_n(x)\nonumber
\end{eqnarray} 
is accompanied by a corresponding expansion of the generating functional~:
\begin{equation} 
Z[j] = Z_2[j] + Z_4[j] + Z_6[j] + \dots
\end{equation} 

Functional integration of the quantum fluctuations around the
classical solution gives rise to the loop expansion. The classical
solution is defined as
\begin{equation} 
\left. \displaystyle\frac{\delta S_2[\varphi,j]}{\delta \varphi_i} 
\right|_{\varphi=\varphi_{\rm cl}}
= 0 \qquad \Rightarrow \qquad \varphi_{\rm cl}[j]
\end{equation} 
and it can be constructed iteratively as a functional of the external
fields $j$. Note that we define $\varphi_{\rm cl}[j]$ through the 
lowest--order Lagrangian $\cL_2(\varphi,j)$ at any order in the
chiral expansion. In this case,
$\varphi_{\rm cl}[j]$ carries precisely the tree structure of $O(p^2)$
allowing for a straightforward chiral counting. This would not be true
any more if we had included higher--order chiral Lagrangians in the
definition of the classical solution.

With a mass--independent regularization method like dimensional
regularization, it is straightforward to compute the degree of
homogeneity of a generic Feynman amplitude as a function of external
momenta and meson masses. This number is called the chiral
dimension $D$ of the amplitude and it characterizes the order of
the low--energy expansion. For a connected amplitude with $L$
loops and with $N_n$ vertices of $O(p^n)$ ($n=$ 2,4,6,\dots), it is
given by (Weinberg 1979)
\begin{equation} 
D = 2L + 2 + \sum_n (n-2) N_n~, \qquad n = 4,6,\ldots \label{eq:DL}
\end{equation} 

For a given amplitude, the chiral dimension
obviously increases with $L$. In order to reproduce the (fixed)
physical dimension of the amplitude, each loop produces a factor $1/F^2$.
Together with the geometric loop factor $(4\pi)^{-2}$, the loop expansion
suggests
\begin{equation} 
4\pi F_\pi = 1.2 \mbox{ GeV} \label{eq:naive}
\end{equation} 
as natural scale of the chiral expansion (Manohar and Georgi
1984). Restricting the domain of applicability of CHPT to momenta 
$|p| \la O(M_K)$, the natural expansion parameter of chiral amplitudes
is therefore expected to be of the order
\begin{equation} 
\displaystyle\frac{M_K^2}{16 \pi^2 F_\pi^2} = 0.18~. \label{eq:omag}
\end{equation} 
As we will see soon, these terms often appear multiplied with chiral
logarithms. Substantial higher--order corrections in the chiral
expansion are therefore to be expected
for chiral $SU(3)$. On the other hand, for $N_f=2$ and for momenta 
$|p| \la O(M_\pi)$ the chiral expansion is expected to converge
considerably faster.

The formula
(\ref{eq:DL}) implies that $D=2$ is only possible for $L=0$: the 
tree--level amplitudes from the Lagrangian $\cL_2$ are then polynomials
of degree 2 in the external momenta and masses. The corresponding
generating functional is given by the classical action:
\begin{equation}
Z_2[j] = \int d^4x \cL_2 (\varphi_{\rm cl}[j],j)~.
\end{equation}

Already at next--to--leading order, the amplitudes are not just
polynomials of degree $D=4$, but they are by definition of the chiral
dimension always ho\-mo\-geneous functions of degree $D$
in external momenta and masses. For $D=4$, we have two types of
contributions: either $L=0$ with $N_4=1$, i.e., exactly one vertex of
$O(p^4)$, or $L=1$ and only vertices of $O(p^2)$ (which, as formula
(\ref{eq:DL}) demonstrates, do not modify the chiral
dimension). Explicitly, the complete generating functional of $O(p^4)$
consists of

\begin{center}
\begin{tabular}{lll}
$L=0$\qquad & \mbox{     }\qquad $\int d^4x \cL_4(\varphi_{\rm cl}[j],j)$ & 
\mbox{     } \qquad chiral action of $O(p^4)$\\
      & \mbox{     } \qquad $Z_{\rm WZW}[\varphi_{\rm cl}[j],v,a]$ & 
\mbox{     } \qquad chiral anomaly \\
$L=1$\qquad & \mbox{     } \qquad $Z_4^{(L=1)}[j]$ & \mbox{     } \qquad
one--loop functional 
\end{tabular}
\end{center}

\noindent
In addition to the Wess--Zumino--Witten functional $Z_{\rm WZW}$
(Wess and Zumino 
1971; Witten 1983) accounting for the chiral anomaly, the $L=0$ part
involves the general chiral Lagrangian $\cL_4$ with 10 LECs
(Gasser and Leutwyler 1985):
\begin{eqnarray} 
{\cal L}_4 & = & L_1 \langle D_\mu U^\dagger D^\mu U\rangle^2 +
                 L_2 \langle D_\mu U^\dagger D_\nu U\rangle
                     \langle D^\mu U^\dagger D^\nu U\rangle \nonumber \\*
& & + L_3 \langle D_\mu U^\dagger D^\mu U D_\nu U^\dagger D^\nu U\rangle +
    L_4 \langle D_\mu U^\dagger D^\mu U\rangle \langle \chi^\dagger U +
    \chi U^\dagger \rangle  \nonumber \\*
& & +L_5 \langle D_\mu U^\dagger  D^\mu U(\chi^\dagger  U + U^\dagger 
\chi)\rangle
    +
    L_6 \langle \chi^\dagger  U + \chi U^\dagger  \rangle^2 +
    L_7 \langle \chi^\dagger  U - \chi U^\dagger  \rangle^2  \nonumber \\*
& & + L_8 \langle \chi^\dagger  U \chi^\dagger  U +
 \chi U^\dagger  \chi U^\dagger \rangle
    -i L_9 \langle F_R^{\mu\nu} D_\mu U D_\nu U^\dagger  +
      F_L^{\mu\nu} D_\mu U^\dagger  D_\nu U \rangle \nonumber \\*
& & + L_{10} \langle U^\dagger  F_R^{\mu\nu} U F_{L\mu\nu}\rangle +
    \mbox{   2 contact terms} = \sum_i L_i P_i
\label{eq:L4}
\end{eqnarray} 
where $F_R^{\mu\nu}$, $F_L^{\mu\nu}$ are field strength tensors
associated with the external gauge fields.
This is the most general Lorentz invariant Lagrangian of $O(p^4)$ with
(local) chiral symmetry, parity and charge conjugation. 

The one--loop functional can be  written in closed form as
\begin{equation} 
Z_4^{(L=1)}[j] = \displaystyle\frac{i}{2} \ln \det D_2 = 
\displaystyle\frac{i}{2} \mbox{ Tr }\ln D_2 \label{eq:Loop1}
\end{equation} 
in terms of the determinant of a differential operator associated with
the Lagrangian $\cL_2$. In accordance with general theorems of
renormalization theory (e.g., Collins 1984), its divergent part takes 
the form of a local
action with all the symmetries of $\cL_2$ and thus of QCD.
Since the chiral dimension of this divergence action is
4, it must be of the form (\ref{eq:L4}) with divergent coefficients:
\begin{eqnarray} 
\cL^{(L=1)}_{4,{\rm div}}& =& - \Lambda(\mu) \sum_i \Gamma_i P_i \\
\Lambda(\mu) &=& \displaystyle\frac{\mu^{d-4}}{(4\pi)^2} 
\left\{\displaystyle\frac{1}{d-4} - \displaystyle\frac{1}{2}
[\ln 4\pi + 1 + \Gamma'(1)]\right\}\nonumber 
\end{eqnarray} 
with the conventions of Gasser and Leutwyler (1985) for
$\overline{{\rm MS}}$. The coefficients $\Gamma_i$ are listed in Table
\ref{tab:Li}.

\renewcommand{\arraystretch}{1.1}
\begin{table}[t]
\begin{center}
\caption{Phenomenological values of the renormalized LECs
$L^r_i(M_\rho)$, taken from Bijnens et al. (1995), and $\beta$ functions
 $\Gamma_i$ for these coupling constants.} \label{tab:Li}
\vspace{.5cm}
\begin{tabular}{|c||r|l|r|}  \hline
i & $L^r_i(M_\rho) \times 10^3$ & \hspace*{.5cm} source & $\Gamma_i$ \\ \hline
  1  & 0.4 $\pm$ 0.3 & $K_{e4},\pi\pi\rightarrow\pi\pi$ & 3/32  \\
  2  & 1.35 $\pm$ 0.3 &  $K_{e4},\pi\pi\rightarrow\pi\pi$&  3/16  \\
  3  & $-$3.5 $\pm$ 1.1 &$K_{e4},\pi\pi\rightarrow\pi\pi$&  0     \\
  4  & $-$0.3 $\pm$ 0.5 & Zweig rule &  1/8  \\
  5  & 1.4 $\pm$ 0.5  & $F_K:F_\pi$ & 3/8  \\
  6  & $-$0.2 $\pm$ 0.3 & Zweig rule &  11/144  \\
  7  & $-$0.4 $\pm$ 0.2 &Gell-Mann--Okubo,$L_5,L_8$ & 0             \\
  8  & 0.9 $\pm$ 0.3 & \small{$M_{K^0}-M_{K^+},L_5,$}&
5/48 \\
     &               &   \small{ $(2m_s-m_u-m_d):(m_d-m_u)$}       & \\
 9  & 6.9 $\pm$ 0.7 & $\langle r^2\rangle^\pi_V$ & 1/4  \\
 10  & $-$5.5 $\pm$ 0.7& $\pi \rightarrow e \nu\gamma$  &  $-$ 1/4  \\
\hline
\end{tabular}
\end{center}
\end{table}
 
Renormalization to $O(p^4)$ proceeds by decomposing
\begin{equation} 
L_i = L_i^r(\mu) + \Gamma_i \Lambda(\mu)
\end{equation} 
such that 
\begin{equation} 
Z_4 - Z_{\rm WZW} = Z_4^{(L=1)} + \int d^4x \cL_4(L_i) = Z^{(L=1)}_{4,{\rm fin}}
(\mu) + \int d^4x \cL_4(L_i^r(\mu))
\end{equation}
is finite and independent of the arbitrary scale $\mu$. The generating
functional and therefore the amplitudes depend on
scale dependent LECs that obey the renormalization group equations
\begin{equation} 
L_i^r(\mu_2) = L_i^r(\mu_1) + \displaystyle\frac{\Gamma_i}{(4\pi)^2} \ln
\displaystyle\frac{\mu_1}{\mu_2}~.\label{eq:RGE}
\end{equation} 
The current values of these constants come mainly from phenomenology
to $O(p^4)$ and are listed in Table \ref{tab:Li}.

\begin{figure}[t]
\vspace*{1cm}
\centerline{\epsfig{file=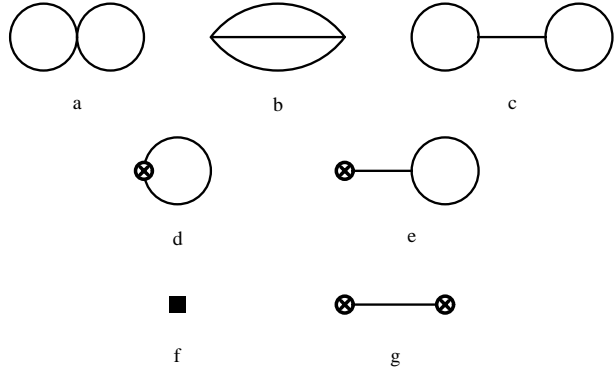,height=5cm}}
\caption{Skeleton diagrams of $O(p^6)$. Normal vertices are from
$\cL_2$, crossed circles denote vertices from $\cL_4$ and the square in 
diagram f stands for a vertex from $\cL_6$.
The propagators and vertices carry the full tree structure associated
with the lowest--order Lagrangian $\cL_2$.}
\label{fig:2loop}
\end{figure}

Many recent investigations in CHPT have included effects of $O(p^6)$ 
(see below for a discussion of elastic $\pi\pi$ scattering). 
The following contributions
are also shown pictorially in Fig.~\ref{fig:2loop}:
\begin{eqnarray*} 
D = 6 : &\qquad L = 0, &\qquad N_6 = 1 \\*
        &\qquad L = 0, &\qquad N_4 = 2 \\
	&\qquad L = 1, &\qquad N_4 = 1 \\
	&\qquad L = 2 & 
\end{eqnarray*} 

Unlike for the one--loop functional (\ref{eq:Loop1}), no simple closed 
form for the two--loop functional $Z_6^{(L=2)}[j]$ is known. General 
theorems of renormalization theory guarantee that
\begin{itemize} 
\item the sum of the irreducible loop diagrams a, b, d in
Fig.~\ref{fig:2loop} is free of subdivergences, and that
\item the sum of the one--particle--reducible diagrams c,
e, g is finite and scale independent (at least for the form of
$\cL_4$ given in (\ref{eq:L4})).
\end{itemize} 
As a consequence,  $Z_{\rm 6,div}^{(L=2)}$ is again a local action with all
the symmetries of $\cL_2$ and the corresponding divergence Lagrangian
is of the general form $\cL_6$ with divergent coefficients. For
general $N_f$, this Lagrangian has 115 terms (Bijnens et al. 1998b),
112 measurable LECs and three contact terms. For $N_f=3$, this
Lagrangian was first written down by Fearing and Scherer (1996) but 
some of their terms are redundant.

How does renormalization at $O(p^6)$ work in practice? To simplify the
discussion, we consider chiral $SU(2)$ with a single
mass scale $M$ (the pion mass at lowest order). 
The LECs in chiral $SU(2)$ and their associated $\beta$ functions are 
usually denoted $l_i, \gamma_i$
(Gasser and Leutwyler 1984). Since the divergences
occur only in polynomials in the external momenta and masses, we
consider a generic dimensionless coefficient $Q$ of such a polynomial,
e.g.,  $m_6$, $f_6$ in the chiral expansions of the pion mass and
decay constant, respectively (B\"urgi 1996; Bijnens et al. 1997):
\begin{eqnarray} 
M_\pi^2&=&M^2\left\{1+m_4 \displaystyle\frac{M^2 }{F^2} 
+m_6\displaystyle\frac{M^4}{F^4}+O(F^{-6})\right\}\\*
F_\pi&=&F\left\{1+f_4 \displaystyle\frac{M^2 }{F^2} 
+f_6\displaystyle\frac{M^4}{F^4}+O(F^{-6})\right\}~.
\end{eqnarray} 
Working from now on in $d$ dimensions, we obtain from the
(irreducible) diagrams a,b,d and f 
\begin{equation} 
Q = Q_{\rm loop} + Q_{\rm tree}
\end{equation} 
with
\begin{equation} 
Q_{\rm loop}(d) = \underbrace{J(0)^2 x(d)}_{\mbox{diagrams a,b}}
 + \underbrace{J(0) \sum_i l_i y_i(d)}_{\mbox{diagram d}}
\end{equation} 
$$
J(0)=\displaystyle\frac{1}{i} \int \displaystyle\frac{d^dk}{(2\pi)^d}
\displaystyle\frac{1}{(k^2-M^2)^2}~.
$$
The coefficients $x(d),y_i(d)$ are expanded to $O(\omega^2)$ in
$\omega = \frac{1}{2} (d-4)$:
\begin{eqnarray} 
x(d) &=& x_0 + x_1 \omega + x_2 \omega^2 + O(\omega^3)  \\
y_i(d) &=& y_{i0} + y_{i1} \omega + y_{i2} \omega^2 + O(\omega^3)~.
\nonumber
\end{eqnarray} 
Likewise, for $J(0)$ and the (unrenormalized) $l_i$ we perform a
Laurent expansion in $\omega$:
\begin{eqnarray} 
\label{eq:Jli}
J(0) &=& \displaystyle\frac{M^{2\omega}\Gamma(-\omega)}{(4\pi)^{2+\omega}} =
\displaystyle\frac{(c\mu)^{2\omega}}{(4\pi)^2} \left( 
\displaystyle\frac{M}{c\mu}\right)^{2\omega}
\displaystyle\frac{\Gamma(-\omega)}{(4\pi)^\omega}\\
& =& \displaystyle\frac{(c\mu)^{2\omega}}{(4\pi)^2} \left[ - 
\displaystyle\frac{1}{\omega} +
b(M/\mu) + a(M/\mu)\omega + O(\omega^2)\right]  \nonumber \\
l_i & = & \displaystyle\frac{(c\mu)^{2\omega}}{(4\pi)^2} \left[ 
\displaystyle\frac{\gamma_i}{2\omega}
+ \beta_i(\mu) + \alpha_i(\mu) \omega + O(\omega^2)\right]~.
\end{eqnarray} 
In the $\overline{MS}$ scheme with
\begin{equation} 
2 \ln c = -1 - \ln 4\pi - \Gamma'(1)
\end{equation} 
one gets
\begin{equation} 
b(M/\mu) = - 2 \ln \displaystyle\frac{M}{\mu} - 1
\end{equation} 
$$
\beta_i(\mu) = (4\pi)^2 l_i^r(\mu)
$$
where the $l_i^r(\mu)$ are the standard renormalized LECs of Gasser
and Leutwyler (1984).

An important consistency check is due to the absence of nonlocal
divergences of the type
$$
\displaystyle\frac{\ln{M/\mu}}{\omega}
$$
implying (Weinberg 1979)
\begin{equation} 
4 x_0 = \sum_i \gamma_i y_{i0}~.\label{eq:WR}
\end{equation} 
For $SU(N_f)$, there are 115 such relations between two--loop and
one--loop quantities due to the 115 independent monomials in the
chiral Lagrangian of $O(p^6)$. We have recently verified these
conditions by explicit calculation (Bijnens et al. 1998b).

With the summation convention for $i$ implied, the
complete loop contribution
\begin{eqnarray} 
Q_{\rm loop} &=& \displaystyle\frac{\mu^{4\omega}}{(4\pi)^4} 
\left\{ - \displaystyle\frac{x_0}{\omega^2} + 
\displaystyle\frac{[x_1-\beta_i(\mu)y_{i0} -
\displaystyle\frac{1}{2} \gamma_i y_{i1}]}{\omega} \right. \\
&& \mbox{} + x_0 b(M/\mu)^2 + \left[-2x_1 + \beta_i(\mu)y_{i0} + 
\displaystyle\frac{1}{2} \gamma_i y_{i1}\right] b(M/\mu)\nonumber\\
& & \mbox{} + \left. x_2 - \beta_i(\mu) y_{i1} -
\displaystyle\frac{1}{2} \gamma_i y_{i2} - \alpha_i(\mu) y_{i0} 
+ O(\omega) \right\}\nonumber
\end{eqnarray} 
is renormalized by the tree--level contribution from $\cL_6$:
\begin{eqnarray} 
Q_{\rm tree}(d) &=& z(d) \\
&=& \displaystyle\frac{\mu^{4\omega}}{(4\pi)^4}
\left\{\displaystyle\frac{x_0}{\omega^2} - 
\displaystyle\frac{[x_1 - \beta_i(\mu)y_{i0} - 
\displaystyle\frac{1}{2} \gamma_i y_{i1}]}
{\omega} + (4\pi)^4 z^r(\mu) + O(\omega) \right\}\nonumber
\end{eqnarray} 
where $z$ is the appropriate combination of (unrenormalized) LECs of
$O(p^6)$. The total contribution from diagrams a,b,d,f is now finite 
and scale independent:
\begin{eqnarray} 
Q &=& \lim_{d \rightarrow 4} \left[ Q_{\rm loop}(d) + Q_{\rm
tree}(d)\right] \\
&=&\displaystyle\frac{1}{(4\pi)^4} \left\{ x_0 \left[ 1 + 2\ln 
\displaystyle\frac{M}{\mu}
\right]^2 + \left[2x_1 - \displaystyle\frac{1}{2} \gamma_i y_{i1} -
(4\pi)^2 l_i^r(\mu) y_{i0}\right]
\left( 1 + 2 \ln \displaystyle\frac{M}{\mu}\right)  \right.\nonumber  \\
& & \mbox{} + \left. x_2 - \displaystyle\frac{1}{2} \gamma_i y_{i2} - 
(4\pi)^2 l_i^r(\mu) y_{i1} + (4\pi)^4 \widehat z^r(\mu)\right\} \nonumber
\end{eqnarray} 
in terms of a redefined\footnote{This process independent
(Bijnens et al. 1998b) redefinition absorbs the
redundant expansion coefficients $\alpha_i(\mu)$.} combination 
$\widehat z^r(\mu)$ of LECs,
\begin{equation} 
\widehat z^r(\mu) = z^r(\mu) - \displaystyle\frac{\alpha_i(\mu) y_{i0}}
{(4\pi)^4}
\end{equation} 
that obeys the renormalization group equation
\begin{equation} 
\mu \displaystyle\frac{d\widehat z^r(\mu)}{d\mu} = 
\displaystyle\frac{2}{(4\pi)^4}
[2x_1 - (4\pi)^2 l_i^r(\mu) y_{i0} - \gamma_i y_{i1}]~.
\end{equation} 

\vspace{.5cm}
\noindent
Remarks: \begin{enumerate} 
\item[i.] Weinberg's relation (\ref{eq:WR}) implies that the
coefficient of the leading chiral log $\ln^2{M/\mu}$
can be extracted from a one--loop calculation (cf. Kazakov 1988). 
\item[ii.] There are in general additional finite contributions
(including chiral logs) from the reducible diagrams c,e,g of 
Fig.~\ref{fig:2loop}.
\end{enumerate} 

In Table \ref{tab:2loop}, I list the complete two--loop
calculations that have been performed up to now. The first five
entries are for chiral $SU(2)$, the last two for $N_f=3$. 

\renewcommand{\arraystretch}{1.1}
\begin{table}[t]
\begin{center}
\caption{Complete calculations to $O(p^6)$ in standard CHPT.} 
\label{tab:2loop}
\vspace*{1cm}
\begin{tabular} {llll}
$\gamma \gamma \to \pi^0 \pi^0$ & \mbox{            }& \mbox{       }
  &  Bellucci et al. (1994) \\
$\gamma \gamma \to \pi^+ \pi^-$ & \mbox{            }& & B\"urgi
(1996)  \\
$\pi \to l \nu_l \gamma$ &\mbox{            }& & Bijnens and Talavera
(1997)  \\
$\pi \pi \to \pi \pi$ &\mbox{            }& & Bijnens et al. (1996, 1997) \\
$\pi$ form factors &\mbox{            }& & Bijnens et al. (1998a)\\
\hline 
$VV$, $AA$ &\mbox{            }& & Golowich and Kambor (1995,
1997) \\
form factors  &\mbox{            }& & Post and Schilcher (1997) \\
\hline
\end{tabular} 
\end{center}
\end{table}

\subsection{Light Quark Masses}

In the framework of standard CHPT, the (current) quark masses $m_q$ 
always appear in the combination $m_q B$ in chiral amplitudes.
Without additional information on $B$ through the
quark condensate [cf. Eq.~(\ref{eq:FB})], one can 
only extract ratios of quark masses from CHPT amplitudes.

\begin{figure}[t]
\vspace*{1cm}
\centerline{\epsfig{file=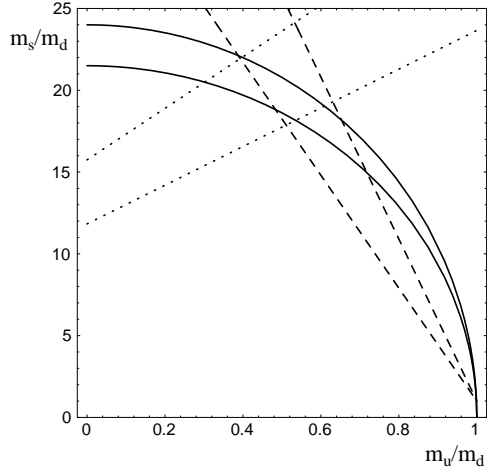,height=6cm}}
\caption{First quadrant of Leutwyler's ellipse 
for $Q=24$ (upper curve) and $Q=21.5$ (lower curve). The dotted lines
correspond to $\Theta_{\eta \eta'}=-15^0$ (upper line) and $-25^0$
(lower line) for the $\eta-\eta'$ mixing angle. The bounds
defined by the two dashed lines come from baryon
mass splittings, $\rho - \omega$ mixing and $\Gamma(\psi'\to
\psi \pi^0)/\Gamma(\psi'\to \psi \eta)$ (Leutwyler 1996b, 1996c) for
the ratio $R=(m_s - \hat m)/(m_d - m_u)$ ($35 \leq R \leq 50$).}
\label{fig:ellipse}
\end{figure}

The lowest--order mass formulas
(\ref{eq:ratios}) together with Dashen's theorem on the lowest--order 
electromagnetic contributions to the meson masses (Dashen 1969) lead 
to the ratios (Weinberg 1977)
\begin{equation} 
\displaystyle\frac{m_u}{m_d} = 0.55~, \qquad \qquad
\displaystyle\frac{m_s}{m_d} = 20.1 ~. \label{eq:mratio}
\end{equation}
Generalized CHPT, on the other hand, does not fix these ratios even at
lowest order but only yields bounds (Fuchs et al. 1990), e.g.,
\begin{equation} 
6 \le r:=\displaystyle\frac{m_s}{\hat{m}} \le
r_2:=\displaystyle\frac{2 M_K^2}{M_\pi^2}-1 \simeq 26 
\label{eq:rler2}
\end{equation} 
with $2 \hat{m}:= m_u+m_d$.
The ratios (\ref{eq:mratio}) receive higher--order corrections. The
most important ones are corrections of $O(p^4) = O(m_q^2)$ and 
$O(e^2 m_s)$. Gasser and Leutwyler (1985)
found that to $O(p^4)$ the ratios 
\begin{eqnarray} 
\displaystyle\frac{M_K^2}{M_\pi^2} &=& \displaystyle\frac{m_s + 
\hat{m}}{m_u + m_d} [1 + \Delta_M +
O(m_s^2)] \label{eq:MKpi} \\
\displaystyle\frac{(M_{K^0}^2 - M_{K^+}^2)_{\rm QCD}}{M_K^2 - M_\pi^2} &=&
\displaystyle\frac{m_d - m_u}{m_s - \hat{m}} [1 + \Delta_M + O(m_s^2)]
\end{eqnarray} 
depend on the same correction $\Delta_M$ of $O(m_s)$.
The ratio of these two ratios is therefore independent of $\Delta_M$
and it determines the quantity
\begin{equation} 
Q^2 := \displaystyle\frac{m_s^2 - \hat{m}^2}{m_d^2 - m_u^2}~. 
\label{eq:Qratio}
\end{equation} 
Without higher--order electromagnetic corrections for the meson masses,
$$
Q = Q_D = 24.2 ~,
$$
but those corrections reduce $Q$ by up to 10$\%$ (Donoghue et
al. 1993; Bijnens 1993; Duncan et al. 1996; Kambor et al. 1996;
Anisovich and Leutwyler 1996; Leutwyler 1996a; Baur and Urech 1996;
Bijnens and Prades 1997; Moussallam 1997).
Plotting $m_s/m_d$ versus $m_u/m_d$ leads to an ellipse 
(Leutwyler 1990). In Fig.~\ref{fig:ellipse}, the relevant quadrant of
the ellipse is shown for $Q = 24$ (upper curve) and $Q = 21.5$ 
(lower curve).

Kaplan and Manohar (1986) pointed out that due to an accidental
symmetry of $\cL_2+\cL_4$ the separate mass ratios 
$m_u/m_d$ and $m_s/m_d$ cannot be calculated to $O(p^4)$ from S--matrix
elements or $V,A$ Green functions only. Some additional input is
needed like resonance saturation (for (pseudo-)scalar Green
functions), large--$N_c$ expansion, baryon mass splittings, etc.
Some of those constraints are also shown in Fig.~\ref{fig:ellipse}. 
A careful analysis of all available information on the mass ratios
was performed by Leutwyler (1996b, 1996c), with the main
conclusion that the quark mass ratios change rather
little from $O(p^2)$ to $O(p^4)$. In Table \ref{tab:ratios}, I compare
the so--called current algebra mass ratios of $O(p^2)$ with the ratios 
including $O(p^4)$ corrections, taken from Leutwyler (1996b, 1996c). 
The errors are Leutwyler's
estimates of the theoretical uncertainties as of 1996. Although 
theoretical errors are always open to debate, the overall stability 
of the quark mass ratios is evident.
 
\renewcommand{\arraystretch}{1.5}
\begin{table}[t]
\begin{center}
\caption{Quark mass ratios at $O(p^2)$ (Weinberg 1977) and to $O(p^4)$
(Leutwyler 1996b, 1996c).} \label{tab:ratios}
\vspace{.5cm}
\begin{tabular}{|c|ccc|}  \hline
\mbox{   } &\mbox{   } $m_u/m_d$ \mbox{   }
 &\mbox{   } $m_s/m_d$ \mbox{   } & 
\mbox{   } $m_s/\hat m$ \mbox{   } \\
\hline
\mbox{   } $O(p^2)$ \mbox{   } & 0.55 & 20.1 &25.9  \\
\mbox{   } $O(p^4)$ \mbox{   } &\mbox{   } 0.55 $\pm$ 0.04 \mbox{   }
 &\mbox{   } 18.9 $\pm$ 0.8 \mbox{   }
&\mbox{   } 24.4 $\pm$ 1.5\mbox{   } \\
\hline
\end{tabular}
\end{center}
\end{table}

Let me now turn to the absolute values of the light quark masses. 
Until recently, the results from QCD sum rules (de Rafael 1998 and
references therein) tended to be systematically higher than the quark
masses from lattice QCD. Some lattice determinations were actually
in conflict with rigorous lower bounds on the quark masses (Lellouch
et al. 1997). Recent progress in lattice QCD  (e.g., L\"uscher 1997) 
has led to a general increase of the (quenched) lattice values. Table
\ref{tab:mq} contains the most recent determinations of both 
$\hat m$ and $m_s$ that I am aware of. Judging only on the basis of the
entries in Table \ref{tab:mq}, sum rule and lattice values for the
quark masses now seem to be compatible with each other. The values are
given at the $\overline{MS}$ scale $\nu=2$ GeV as is customary in 
lattice QCD.

\renewcommand{\arraystretch}{1.5}
\begin{table}[t]
\begin{center}
\caption{Light quark masses in MeV at the $\overline{MS}$ scale
$\nu=2$ GeV. The most recent values from QCD sum rules and 
(quenched) lattice calculations are listed.} \label{tab:mq}
\vspace{.5cm}
\begin{tabular}{|ccc|}  \hline
\mbox{   } $\hat m$ \mbox{   } & \mbox{              }
 &\mbox{   } $m_s$ \mbox{   } \\ 
\hline
\mbox{   } 4.9 $\pm$ 0.9 \mbox{   } & \mbox{   } sum rules \mbox{   }
& \mbox{   } 125 $\pm$ 25  \mbox{   } \\
\mbox{   } Prades (1998) \mbox{   } &\mbox{               }
 &\mbox{   } Jamin (1998) \mbox{   } \\ \hline
\mbox{   } 5.7 $\pm$ 0.1 $\pm$ 0.8 \mbox{   }  &
 \mbox{   } lattice \mbox{   }
& \mbox{   } 130 $\pm$ 2 $\pm$ 18 \mbox{   } \\
\mbox{               } &  \mbox{   } Gim\'enez et al. (1998) 
\mbox{      }  & \mbox{               }  \\
\hline
\end{tabular}
\end{center}
\end{table}

\begin{figure}[t]
\vspace*{1cm}
\centerline{\epsfig{file=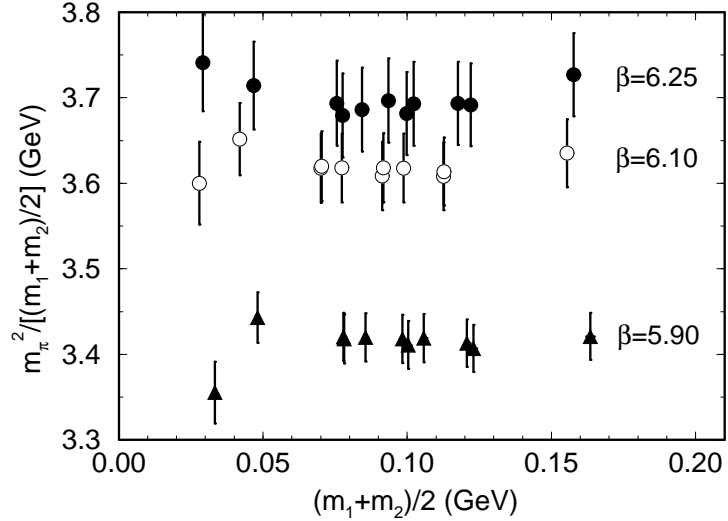,height=7cm}}
\caption{$2 M_\pi^2/(m_1+m_2)$ as a function of $(m_1+m_2)/2$
 (from Aoki et al. 1998).}
\label{fig:aoki}
\end{figure}

Except for chiral logs, the squares of the meson masses are
polynomials in $m_q$. It is remarkable if not puzzling that many
years of lattice studies have not seen any indications for
terms higher than linear in the quark masses. An impressive example
from the high--statistics spectrum calculation of the CP-PACS 
Collaboration (Aoki et al. 1998) is shown in Fig.~\ref{fig:aoki}. 
The ratio $M^2/(m_1+m_2)$ appears to be flat over the whole range of
quark masses accessible in the simulations. The different values of
$\beta$ stand for different lattice spacings but for each $\beta$ the 
ratio is constant to better than 5$\%$. Since lattice calculations
have found evidence for nonlinear quark mass corrections to baryon 
masses (e.g., Aoki et al. 1998), it is difficult to blame
this conspicuous linearity\footnote{Quenching effects are
estimated to be $\sim5\%$ at the lightest $m_q$ presently available on the
lattice (Sharpe 1997; Golterman 1997).}
between $M^2$ and $m_q$ on the limitations
of present--day lattice methods only.

In order to see whether the lattice findings are consistent with CHPT,
I take the $O(p^4)$ result (Gasser and Leutwyler 1985) for $M_K^2$
and vary $m_1=\hat m$, $m_2=m_s$. Since the actual quark masses on the
lattice are still substantially bigger than $\hat m$, the $SU(2)$
result for $M_\pi^2$ cannot be used for this comparison.

Writing $M^2$ instead of $M_K^2$ for general $m_1, m_2$, one
finds
\begin{eqnarray} 
M^2 &=& (m_1+m_2)B\left\{1+\displaystyle\frac{(m_1+2 m_2)B}
{72 \pi^2 F^2}\ln\displaystyle\frac{2(m_1+2 m_2)B}{3 \mu^2}\right.
\label{eq:M2mq}\\*
&+& \left. \displaystyle\frac{8(m_1+m_2)B}{F^2} (2 L_8^r(\mu)-
L_5^r(\mu))+\displaystyle\frac{16(2 m_1+m_2)B}{F^2} (2 L_6^r(\mu)-
L_4^r(\mu))\right\} \nonumber
\end{eqnarray} 
with the scale dependent LECs given in Table \ref{tab:Li}. As can
easily be checked with the help of Eq.~(\ref{eq:RGE}), $M^2$ in 
(\ref{eq:M2mq}) is independent of the arbitrary scale $\mu$ as
it should be.

\begin{figure}[t]
\vspace*{1cm}
\centerline{\epsfig{file=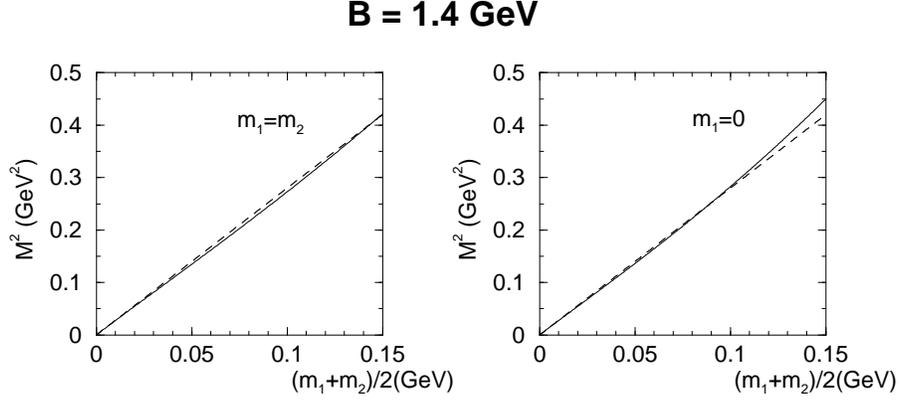,width=12cm}}
\caption{$M^2 (\rm{GeV}^2)$ as function of the average quark mass     
(in GeV) in standard CHPT. The dashed lines are the
lowest--order predictions; the full lines correspond to the
results of $O(p^4)$ in Eq.~(\protect{\ref{eq:M2mq}}).
The two graphs are for $m_1=m_2$ and $m_1=0$, respectively.}
\label{fig:smass}
\end{figure}

Since the $L_i$ are by definition independent of quark masses, it
is legitimate to use the values in Table \ref{tab:Li} also when 
varying $m_1, m_2$. Let me first consider the standard scenario with
$B(\nu=1~{\rm GeV})=1.4$ GeV\footnote{Note that the quark masses in
Fig.~\ref{fig:aoki} correspond to $\nu=2~{\rm GeV}$, however.}
together with the mean values of the $L_i^r(M_\rho)$ in Table
\ref{tab:Li}. In Fig.~\ref{fig:smass}, $M^2$ is plotted as
a function of the average quark mass $(m_1+m_2)/2$ for two extreme 
cases: $m_1=m_2$ or $m_1=0$. The second case with a massless quark 
can of course not be implemented on the lattice. As the figure 
demonstrates, there is little deviation from 
linearity at least up to $M\simeq$ 600 MeV although this deviation is
in general bigger than suggested by Fig.~\ref{fig:aoki} (for the range
of LECs in Table \ref{tab:Li}). 

\begin{figure}[t]
\vspace*{1cm}
\centerline{\epsfig{file=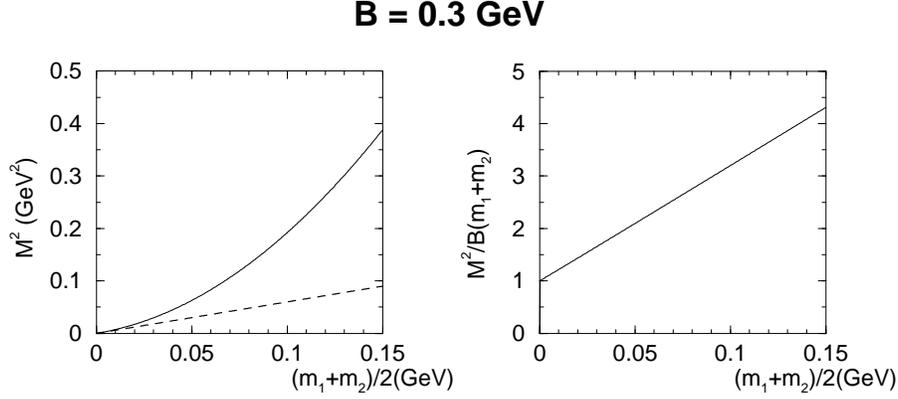,width=12cm}}
\caption{$M^2 (\rm{GeV}^2)$ as function of the average quark mass 
(in GeV) for $B$=0.3 GeV. The notation in the first
graph is as in the previous figure. The second graph shows
the ratio $M^2/B(m_1+m_2)$ from (\protect{\ref{eq:M2mq}}).}
\label{fig:gmass}
\end{figure}

In order to demonstrate that the near--linearity is  specific for
standard CHPT, we now
lower the value of $B$ as suggested by the proponents
of  generalized CHPT. Remember that $B=O(F_\pi)$ is considered to be a
reasonable value in that scenario. To show the dramatic changes
required by a small $B$, I choose an intermediate value 
$B=0.3$ GeV. Of course, in order to obtain the observed 
meson masses, at least some of the LECs have to be scaled up. 
Leaving the signs of the LECs unchanged, 
Eq.~(\ref{eq:M2mq}) requires to scale up $L_8^r$ to obtain realistic 
meson masses for a similar range of quark masses as before. But this 
is precisely the suggestion of generalized CHPT that the LECs associated 
with mass terms in $\cL_4$ may have been underestimated (Stern 1997) by 
standard CHPT. For the following plot, I therefore take $L_8^r(M_\rho)=$ 
20$\cdot10^{-3}$. The two cases considered before ($m_1=m_2$ or
$m_1=0$) are now practically indistinguishable and they lead to a
strong deviation from linearity as exhibited in the first graph of
Fig.~\ref{fig:gmass}. The second graph can be compared 
with the lattice results in Fig.~\ref{fig:aoki}. Please make
sure to compare the scales of the ordinates: whereas the lattice
ratios vary by at most 5$\%$, this ratio would now have to change by more
than a factor of four (!) over the same range of quark masses.

The conclusion of this exercise is straightforward: lattice QCD is
incompatible with a small quark condensate. Unless lattice
simulations for the meson mass spectrum are completely
unreliable, the observed linearity of $M^2$ in the quark masses
favours standard CHPT and excludes values of 
$B$ substantially smaller than the standard value. 

\subsection{Pion--Pion Scattering}

There are several good reasons for studying elastic pion--pion scattering:

\begin{itemize}
\item[i.] The elastic scattering of the lightest hadrons is a
fundamental process for testing CHPT: the only
particles involved are $SU(2)$ pseudo--Goldstone bo\-sons. One may
rightfully expect good convergence of the low--energy expansion
near threshold.
\item[ii.] The behaviour of the scattering amplitude
near threshold is sensitive to the mechanism of
spontaneous chiral symmetry breaking (Stern et al. 1993),
or more precisely, to the size of the quark condensate.
\item[iii.] After a long period without much experimental
activity, there are now good prospects for significant
improvements in the near future. $K_{e4}$ experiments to extract
pion--pion phase shifts due to the final--state interactions of the
pions are already in the analysis stage at Brookhaven (Lowe 1997) or
will start this year at the $\Phi$ factory DA$\Phi$NE in 
Frascati (Baillargeon and Franzini 1995; Lee-Franzini 1997). In
addition, the ambitious DIRAC
experiment (Adeva et al. 1994; Schacher 1997) is being set up at CERN
to measure a combination of $S$--wave scattering lengths through a
study of $\pi^+\pi^-$ bound states.
\end{itemize}

In the isospin limit $m_u=m_d$, the scattering amplitude is
determined by one scalar function $A(s,t,u)$ of the
Mandelstam variables. In terms of this function, one can
construct amplitudes with definite isospin ($I=0,1,2$) in the
$s$--channel. A partial--wave expansion 
gives rise to partial--wave amplitudes $t_l^I(s)$ that are
described by real phase shifts $\delta_l^I(s)$ in the elastic region
$4M_\pi^2 \leq s \leq 16M_\pi^2$ in the usual way:
\begin{equation}
t_l^I(s)=(1-\displaystyle \frac{4 M_\pi^2}{s})^{-1/2} \exp{i
\delta_l^I(s)} \sin{\delta_l^I(s)}~.
\end{equation}
The behaviour of the partial waves near threshold is of the form
\begin{equation}
\Re e\;t_l^I(s)=q^{2l}\{a_l^I +q^2 b_l^I +O(q^4)\}~,
\label{eq:effr}
\end{equation}
with $q$ the center--of--mass momentum.  The quantities $a_l^I$ and
$b_l^I$ are referred to as scattering lengths and slope parameters,
respectively.

The low--energy expansion for $\pi\pi$ scattering has been
carried through to $O(p^6)$ where two--loop diagrams must be
included. Before describing the more recent work, let me recall
the results at lower orders.\\

\begin{center}
\underline{$O(p^2)$ ($L=0$)}
\end{center}

As discussed previously in this lecture, only tree diagrams from
the lowest--order Lagrangian $\cL_2$ contribute at $O(p^2)$. 
The scattering amplitude was first written
down by Weinberg (1966):
\begin{equation}
A_2(s,t,u)=\displaystyle\frac{s-M_\pi^2}{F_\pi^2}~. 
\end{equation}
At the same order in the standard scheme, the quark mass ratios are 
fixed in terms of meson mass ratios, e.g., $r=r_2$ in the notation of 
Eq.~(\ref{eq:rler2}).

In generalized CHPT, some of the terms in $\cL_4$ in the
standard counting appear already at lowest order. Because there
are now more free parameters, the relation $r=r_2$ is replaced
by the bounds (\ref{eq:rler2}). The $\pi\pi$ scattering amplitude
of lowest order in generalized CHPT is (Stern et al. 1993)
\begin{equation}
A_2(s,t,u)=\displaystyle\frac{s-\frac{4}{3}M_\pi^2}{F_\pi^2}
+ \alpha \displaystyle\frac{M_\pi^2}{3F_\pi^2}
\end{equation}
$$
\alpha= 1 + \displaystyle\frac{6(r_2-r)}{r^2-1}~, \qquad
\alpha \ge 1~.
$$
The amplitude is correlated with the quark mass ratio $r$.
Especially the $S$--wave is very sensitive to $\alpha$: the
standard value of $a_0^0=0.16$ for $\alpha=1$ ($r=r_2$) moves to
$a_0^0=0.26$ for a typical value of $\alpha\simeq 2$ ($r\simeq 10$)
in the generalized scenario. As announced before, the $S$--wave
amplitude is indeed a sensitive measure of the quark mass ratios and
thus of the quark condensate. To settle the issue, the lowest--order
amplitude is of course not sufficient.\\

\begin{center}
\underline{$O(p^4)$ ($L\le 1$)}
\end{center}
\nopagebreak
To next--to--leading order, the scattering
amplitude was calculated by Gasser and Leutwyler (1983):
\begin{eqnarray}
F_\pi^4 A_4(s,t,u)&=&c_1 M_\pi^4 +c_2 M_\pi^2 s +c_3 s^2 + c_4 (t-u)^2
\\
&+& F_1(s) +G_1(s,t) +G_1(s,u) \nonumber~.
\end{eqnarray} 
$F_1, G_1$ are standard one--loop functions and the constants
$c_i$ are linear combinations of the LECs $l_i^r(\mu)$ and of
the chiral log $\ln (M_\pi^2/\mu^2)$. It turns out that many
observables are dominated by the chiral logs. This applies for
instance to the $I=0$ $S$--wave scattering length that increases
from 0.16 to 0.20. This relatively big increase of 
25$\%$ makes it necessary to go still one step further in the chiral
expansion. \\

\begin{center}
\underline{$O(p^6)$ ($L\le 2$)}
\end{center}

Two different approaches have been used. In the dispersive
treatment (Knecht et al. 1995), $A(s,t,u)$ was calculated
explicitly up to a crossing symmetric subtraction polynomial
\begin{equation}
[b_1 M_\pi^4 + b_2 M_\pi^2 s +b_3 s^2 +b_4 (t-u)^2]/F_\pi^4
+[b_5 s^3+b_6 s(t-u)^2]/F_\pi^6
\end{equation}
with six dimensionless subtraction constants $b_i$. Including
experimental information from $\pi\pi$ scattering at higher energies,
Knecht et al. (1996) evaluated four of those constants
($b_3$,\dots, $b_6$) from sum rules. The amplitude is given in a
form compatible with generalized CHPT.

The field theoretic calculation involving Feynman diagrams with
$L=0,1,2$ was performed in the standard scheme (Bijnens et
al. 1996, 1997). Of course, the diagrammatic calculation reproduces
the analytically nontrivial part of the dispersive approach. 
To arrive at the final renormalized amplitude, one needs in addition
the following quantities to $O(p^6)$: the pion wave function
renormalization constant (B\"urgi 1996), the pion mass (B\"urgi 1996)
and the pion decay constant (Bijnens et al. 1996, 1997).
Moreover, in the field theoretic approach the previous subtraction
constants are obtained as functions
\begin{equation}
b_i(M_\pi/F_\pi,M_\pi/\mu;l_i^r(\mu),k_i^r(\mu))~,
\end{equation}
where the $k_i^r$ are six combinations of LECs of the $SU(2)$
Lagrangian of $O(p^6)$.

Compared to the dispersive approach, the diagrammatic method offers
the following advantages:
\begin{description}
\item[i.]  The full infrared structure is exhibited to $O(p^6)$. In
particular, the $b_i$ contain chiral logs of the form
$(\ln{M_\pi/\mu})^n$ ($n\le 2$) that are known to be numerically
important, especially for the infrared--dominated parameters $b_1$ and
$b_2$.
\item[ii.]  The explicit dependence on LECs makes
phenomenological determinations of these constants and comparison with
other processes possible. This is especially relevant for determining
$l_1^r$, $l_2^r$ to $O(p^6)$ accuracy (Colangelo et al. 1998).
\item[iii.]  The fully known dependence on the pion mass allows one to
evaluate the amplitude even at unphysical values of the quark mass
(remember that we assume $m_u=m_d$). One possible application is
to confront the CHPT amplitude with lattice calculations
of pion--pion scattering (Colangelo 1997).
\end{description}

In the standard picture, the $\pi\pi$ amplitude depends on four
LECs of $O(p^4)$ and on six combinations of $O(p^6)$
couplings. The latter have been estimated with meson resonance
exchange that is known to account for the dominant features of the
$O(p^4)$ constants (Ecker et al. 1989).  It turns out (Bijnens
et al. 1997)
that the inherent uncertainties of this approximation induce
small (somewhat bigger) uncertainties for the low (higher) partial waves. The
main reason is that the higher partial waves are more sensitive to the
short--distance structure.

However, as the chiral counting suggests, the LECs of $O(p^4)$ are
much more important. Eventually, the $\pi\pi$
amplitude of $O(p^6)$ will lead to a more precise
determination of some of those constants (Colangelo et al. 1998) than
presently available. For the time being, one can investigate the
sensitivity of the amplitude to the $l_i^r$.
In Table \ref{tab:thresh}, some
of the threshold parameters are listed for three sets of the
$l_i^r$ (Bijnens et al. 1997; Ecker 1997): set I is mainly based
on phenomenology to $O(p^4)$ 
(Gasser and Leutwyler 1984; Bijnens et al. 1994), for set II
the $\pi\pi$ $D$--wave scattering lengths to $O(p^6)$ are used
as input to fix $l_1^r$, $l_2^r$, whereas for set III
resonance saturation is assumed for the $l_i^r$ 
renormalized at $\mu=M_\eta$. Although some of the entries in Table
\ref{tab:thresh} are quite sensitive to the choice of the $l_i^r$, 
two points are worth emphasizing:
\begin{itemize}
\item
The $S$--wave threshold parameters are very stable, especially
the $I=0$ scattering length, whereas the higher partial waves are 
more sensitive to the choice of LECs of $O(p^4)$ 
(and also of $O(p^6)$).
\item
The resonance dominance prediction (set III) is in perfect agreement
with the data although the agreement becomes less impressive for 
$\mu > M_\eta$.
\end{itemize}

\begin{table}[ht]
\caption{Threshold parameters in units of $M_{\pi^+}$ for three sets
of LECs $l_i^r$ (Bijnens et al. 1997; Ecker 1997).
The values of $O(p^4)$ correspond to set I. The experimental values
are from Dumbrajs et al. (1983).}
\label{tab:thresh}
\begin{center}
\begin{tabular}{c||c|c|c|c|c|c}
\hline & & & & & & \\ &\mbox{ } $O(p^2)$\mbox{ } & \mbox{ } $O(p^4)$
\mbox{ }& \mbox{ } $O(p^6)$ \mbox{ }&\mbox{ } $O(p^6)$ \mbox{ }&
\mbox{ } $O(p^6)$ \mbox{ } &\mbox{ } experiment \mbox{ } \\ & & & set
I & set II & set III & \\[2pt] \hline 
$ a_0^0$& $0.16$ & $0.20$ &
$0.217$ & $0.206$ & $0.209$ & $0.26\pm0.05$\\

$b_0^0$ & $0.18$ & $0.25$& $0.275$ &$0.249$ & $0.261$ &
$0.25\pm0.03$\\

$2 a_0^0-5 a_0^2$ & $0.55$ & $0.61$ &$0.641$ & $0.634$ & $0.626$ &
$0.66\pm0.05$\\

$-$10 $b_0^2$ & $0.91$ & $0.73$ &$0.72$ & $0.80$ & $0.75$ &
$0.82\pm0.08$\\

10 $a_1^1$ & $0.30$ & $0.37$ & $0.40$ & $0.38$ & $0.37$ &
$0.38\pm0.02$\\

$10^2 a_2^0$ & $0$ & $0.18$ & $0.27$ & input & $0.19$ & $0.17\pm0.03$
\\ \hline
\end{tabular}
\end{center}
\end{table}

\begin{figure}[t]
\vspace*{1cm}
\begin{center}
\mbox{\epsfysize=8cm \epsfbox{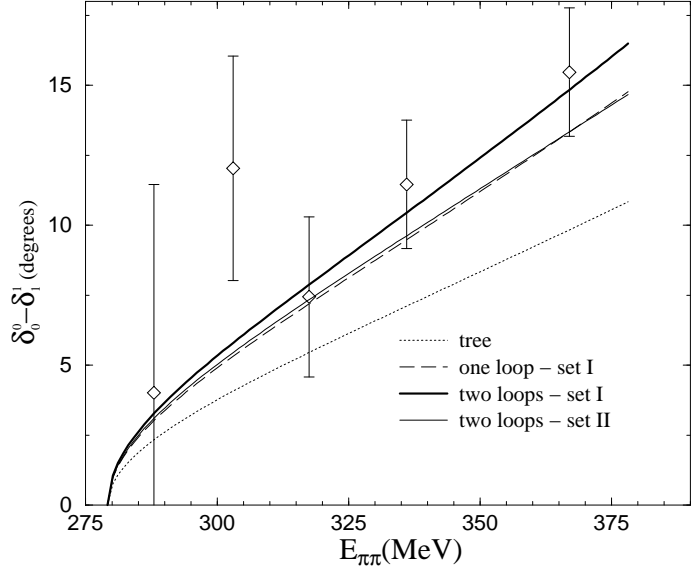} }
\caption{Phase shift difference $\delta_0^0-\delta_1^1$ at $O(p^2)$,
  $O(p^4)$ and $O(p^6)$ (set I and II) from Bijnens et al. (1997).}
\label{fig:d00md11}
\end{center}
\end{figure}

In Fig.~\ref{fig:d00md11}, the phase shift difference
$\delta_0^0 - \delta_1^1$ is plotted as function of the 
center--of--mass energy
and compared with the available low--energy data.  The two--loop phase
shifts describe the $K_{e4}$ data (Rosselet et al. 1977) very well for
both sets I and II, with a small preference for set I. The
curve for set III is not shown in the figure, it lies between those of 
sets I and II.

To conclude this part on $\pi\pi$ scattering,
let me stress the main features:
\begin{itemize}
\item The low--energy expansion converges reasonably well. 
The main uncertainties are not due to the corrections of
$O(p^6)$, but they are related to the LECs of $O(p^4)$. This will 
in turn make a better
determination of those constants possible (Colangelo et al. 1998).
\item Many observables , especially the $S$--wave threshold parameters, 
are infrared dominated
by the chiral logs. This is the reason why the $I=0$ $S$--wave
scattering length is rather insensitive to the LECs of $O(p^4)$. From
the calculations in standard CHPT, a value
\begin{equation}
a_0^0 = 0.21 \div 0.22
\end{equation}
is well established. This will be a crucial test for the standard
framework once the data become more precise. On the basis of 
available experimental information, there is at present no indication
against the standard scenario of chiral symmetry breaking
with a large quark condensate.
\item
Altogether, there is good agreement with the present low--energy data
as both Table \ref{tab:thresh} and Fig.~\ref{fig:d00md11} demonstrate.
\item
Isospin violation and electromagnetic corrections have to be
included. First results are already available (Mei\ss ner et
al. 1997; Knecht and Urech 1997).
\end{itemize}

\section{Baryons and Mesons}

A lot of effort has been spent on the meson--baryon system in
CHPT (e.g., Bernard et al. 1995; Walcher 1998). Nevertheless, the accuracy
achieved is not comparable to the meson sector. Here are some of
the reasons.
\begin{itemize}
\item
The baryons are not Goldstone particles. Therefore, their interactions 
are less constrained by chiral symmetry than for pseudoscalar mesons. 
\item
Due to the fermionic nature of baryons, there are terms of every
positive order in the chiral expansion. In the meson case, only
even orders can contribute.
\item
There are no ``soft'' baryons because the baryon masses stay
finite in the chiral limit. Only baryonic three--momenta
may be soft.
\item
In a manifestly relativistic framework (Gasser et al. 1988), the baryon 
mass destroys the correspondence between loop and chiral
expansion that holds for mesons.
\end{itemize}

In this lecture, I will only consider chiral $SU(2)$, i.e., pions
and nucleons only. Some of the problems mentioned have to do with the
presence of the ``big'' nucleon mass that is in fact comparable
to the scale $4 \pi F_\pi$ of the chiral expansion. 
This comparison suggests a simultaneous expansion in
$$
\displaystyle\frac{\vec{p}}{4\pi F} \qquad \mbox{and} \qquad 
\displaystyle\frac{\vec{p}}{m}
$$
where $\vec{p}$ is a small three--momentum and $m$ is the
nucleon mass in the chiral limit. On the other hand, there is an
essential difference between $F$ and $m$: whereas $F$ appears only 
in vertices, the nucleon mass enters via the nucleon propagator.
To arrive at a simultaneous expansion, one therefore has to shift
$m$ from the propagator to the vertices of some effective
Lagrangian. That is precisely the procedure of
heavy baryon CHPT (Jenkins and Manohar 1991; Bernard et
al. 1992), in close analogy to heavy quark effective theory. 

\subsection{Heavy Baryon Chiral Perturbation Theory}

The main idea of heavy baryon CHPT is to decompose the nucleon
field into ``light'' and ``heavy''
components. In fact, the light components will be
massless in the chiral limit. The heavy components are then
integrated out not unlike other heavy degrees of freedom.
This decomposition is necessarily 
frame dependent but it does achieve the required goal: at the
end, we have an effective chiral Lagrangian with only light degrees
of freedom where the nucleon mass 
appears only in inverse powers in higher--order terms of
this Lagrangian. 

Since the derivation of the effective Lagrangian of heavy baryon
CHPT is rather involved, I will exemplify the method only for
the trivial case of a free nucleon with Lagrangian
\begin{equation}
\cL_0 = \overline{\Psi}(i\dslash - m)\Psi~.
\end{equation}
In terms of a time--like unit four--vector $v$ (velocity), one 
introduces projectors $P_v^\pm = \frac{1}{2} (1 \pm\not\!v)$. 
In the rest system with $v=(1,0,0,0)$, for instance, the $P_v^\pm$ 
project on upper and lower components of the Dirac field in the standard
representation of $\gamma$ matrices. With these projectors, one
defines (Georgi 1990) velocity--dependent fields $N_v$, $H_v$:
\begin{eqnarray}
N_v(x) &=& \exp[i m v \cdot x] P_v^+ \Psi(x) \label{eq:vdf} \\
H_v(x) &=& \exp[i m v \cdot x] P_v^- \Psi(x) \nonumber ~.
\end{eqnarray}
The Dirac Lagrangian is now rewritten in terms of these fields:
\begin{eqnarray}
\cL_0 &=& \overline{(N_v+H_v)}e^{im v\cdot x}(i\dslash - m)
e^{-im v\cdot x}(N_v+H_v) \\
&=& \overline{N_v}i v \cdot \partial N_v - 
\overline{H_v}(i v \cdot \partial +2 m) H_v + {\rm ~mixed ~terms~.}
\nonumber 
\end{eqnarray}
After integrating out the heavy components $H_v$ in the
functional integral with the fully relativistic pion--nucleon 
Lagrangian (Gasser et al. 1988), one
arrives indeed at an effective chiral Lagrangian for the field $N_v$
(and pions) only, with a massless propagator 
\begin{equation} 
\displaystyle\frac{i P_v^+}{v \cdot k + i\varepsilon}~.
\label{eq:Nvprop}
\end{equation} 
At every order except the leading one, $O(p)$, this 
Lagrangian consists of two pieces: the first one is the usual
chiral Lagrangian of $O(p^n)$ with a priori unknown LECs. The
second part comes from the expansion in $1/m$ and it is completely given
in terms of LECs of lower than $n$-th order. Since the only
nucleon field in this Lagrangian is $N_v$ with a massless
propagator, there is a straightforward analogue to chiral power
counting in the meson sector given by formula (\ref{eq:DL}). For a
connected $L$--loop amplitude with $E_B$ external baryon lines and
$N_{n,n_B}$ vertices of chiral dimension $n$ (with
$n_B$ baryon lines at the vertex), the analogue of (\ref{eq:DL}) is
(Weinberg 1990, 1991)
\begin{equation}
D=2 L +2 - \displaystyle\frac{E_B}{2} + \sum_{n,n_B}(n - 2 +
\displaystyle\frac{n_B}{2}) N_{n,n_B}~.\label{eq:DEB}
\end{equation}
However, as we will discuss later on in connection with 
nucleon--nucleon scattering, this formula is misleading for 
$E_B \ge 4$. On the other hand, no problems arise for the case of one 
incoming and one outgoing nucleon ($E_B=2$) where
\begin{equation}
D=2 L + 1  + \sum_n\left[(n - 2)N_{n,0} + (n-1) N_{n,2}
\right] \ge 2 L + 1 ~.\label{eq:DsN}
\end{equation}
This formula is the basis for a systematic low--energy expansion
for single--nucleon processes, i.e., for processes of the type
$\pi N \to \pi \ldots \pi N$, $\gamma N \to \pi \ldots \pi N$, 
$l \,N \to l\,\pi 
\ldots \pi N$ (including nucleon form factors), 
$\nu_l N \to l\, \pi \ldots \pi N$.
The corresponding effective chiral Lagrangian is completely known
to $O(p^3)$ (Bernard et al. 1992; Ecker and Moj\v zi\v s 1996;
Fettes et al. 1998) including the full renormalization at $O(p^3)$
(Ecker 1994):
\begin{eqnarray} 
 \cL_{\pi N}&=& \cL_{\pi N}^{(1)}+ \cL_{\pi N}^{(2)}+
 \cL_{\pi N}^{(3)}+\dots \label{eq:LpiN}\\
 \cL_{\pi N}^{(1)} &=& \overline{N_v}(iv \cdot \nabla + g_A 
S \cdot u) N_v \nonumber
\end{eqnarray} 
$$
u_\mu=i(u^\dagger \partial_\mu u - u \partial_\mu u^\dagger)+
{\rm~external ~gauge ~fields}~, \qquad S^\mu = i \gamma_5 
\sigma^{\mu\nu} v_\nu/2 
$$
with a chiral and gauge covariant derivative $\nabla$ and with 
$g_A$ the axial--vector coupling constant in the chiral limit.\\

\noindent
Two remarks are in order at this point.
\begin{itemize}
\item Since the Lagrangian (\ref{eq:LpiN}) was derived from a fully 
relativistic Lagrangian it defines a Lorentz invariant quantum field 
theory although it depends explicitly on the arbitrary frame vector 
$v$ (Ecker and Moj\v zi\v s 1996). Reparametrization invariance
(Luke and Manohar 1992) is automatically fulfilled.
\item The transformation from the original Dirac field $\Psi$ to the
velocity--dependent field $N_v$ leads to an unconventional wave
function renormalization of $N_v$ that is in general
momentum dependent (Ecker and Moj\v zi\v s 1997).
\end{itemize}

Since the theory is Lorentz invariant it must always be possible
to express the final amplitudes in a manifestly relativistic
form. Of course, this will only be true up to the given order in
the chiral expansion one is considering. The general procedure
of heavy baryon CHPT for single--nucleon processes can then be
summarized as follows.

\begin{itemize}
\item[i.] Calculate the heavy baryon amplitudes to a given chiral 
order with the Lagrangian (\ref{eq:LpiN}) in a frame defined by the 
velocity vector $v$.
\item[ii.] Relate those amplitudes to their relativistic counterparts
which are independent of $v$ to the order considered. For the
special example of the initial nucleon rest frame with $v=
p_{\rm in}/m_N$, the translation is given in Table \ref{tab:INRF}
(Ecker and Moj\v zi\v s 1997).
\item[iii.] Apply wave function renormalization for the external
nucleons.
\end{itemize}

\renewcommand{\arraystretch}{2.}
\begin{table}[t]
\caption{Relations between relativistic covariants and the corresponding
quantities in the initial nucleon rest frame ($v=p_{\rm in}/m_N$,
$q=p_{\rm out}-p_{\rm in}$, $t=q^2$) with 
$\overline{u}(p_{\rm out})\Gamma u(p_{\rm in}) = 
\overline{u}(p_{\rm out})P^+_v \widehat \Gamma P^+_v u(p_{\rm in})$.}
\label{tab:INRF}
$$
\begin{tabular}{|c|c|} \hline
$\Gamma$ & $\widehat \Gamma$ \\ \hline
$1$ & $1$ \\
$\gamma_5$ & $\displaystyle\frac{q \cdot S}{m_N(1 - t/4m^2_N)}$ \\
$\gamma^\mu$ & $\left( 1 - t/4m^2_N\right)^{-1}
\left( v^\mu + \displaystyle\frac{q^\mu}{2m_N} + 
\displaystyle\frac{i}{m_N} \varepsilon^{\mu\nu\rho\sigma}
q_\nu v_\rho S_\sigma\right)$ \\
$\gamma^\mu \gamma_5$ & $2S^\mu - 
\displaystyle\frac{q \cdot S}{m_N(1 - t/4m^2_N)} v^\mu$ \\
$\sigma^{\mu\nu}$ & $2 \varepsilon^{\mu\nu\rho\sigma} v_\rho S_\sigma +
\displaystyle\frac{1}{2m_N(1 - t/4m^2_N)}
\{ i(q^\mu v^\nu - q^\nu v^\mu) + 2(v^\mu 
\varepsilon^{\nu\lambda\rho\sigma}-
v^\nu \varepsilon^{\mu\lambda\rho\sigma}) q_\lambda v_\rho S_\sigma\}$ \\
& \\ 
\hline
\end{tabular}
$$
\end{table}

As an application of this procedure, I will now discuss
elastic pion--nucleon scattering to $O(p^3)$ in the low--energy 
expansion. For other applications of CHPT to 
single--nucleon processes, I refer to the available reviews
(Bernard et al. 1995; Ecker 1995) and 
conference proceedings (Bernstein and Holstein 1995; Walcher 1998).

\subsection{Pion--Nucleon Scattering}

Elastic $\pi N$ scattering is maybe the most intensively studied
process of hadron physics, with a long history both in theory
and experiment (e.g., H\"ohler 1983). 
The systematic CHPT approach is however comparatively 
new (Gasser et al. 1988). I am going to review here the first
complete calculation to $O(p^3)$ by Moj\v zi\v s (1998).
As for $\pi\pi$ scattering, isospin symmetry is assumed.

A comparison with elastic $\pi\pi$ scattering displays the difficulties 
of the $\pi N$ analysis. Although calculations have
been performed to next--to--next--to--leading
order for both processes, this is only $O(p^3)$ for $\pi N$ compared 
to $O(p^6)$ for $\pi\pi$. Of course, this is due to the fact that,
unlike for mesons only, every
integer order can contribute to the low--energy expansion in the
meson--baryon sector. The difference in accuracy also manifests
itself in the number of LECs: the numbers are again comparable
despite the difference in chiral orders.
Finally, while we now know the $\pi\pi$ amplitude to
two--loop accuracy the $\pi N$ amplitude is still not completely
known even at the one--loop level as long as the $p^4$ amplitude
has not been calculated.

The amplitude for pion--nucleon scattering
\begin{equation}
\pi^a(q_1) + N(p_1) \to \pi^b(q_2) + N(p_2)
\end{equation}
can be expressed in terms of four invariant amplitudes $D^\pm$,
$B^\pm$:
\begin{eqnarray}
T_{ab} &=& T^+ \delta_{ab} - T^- i \varepsilon_{abc} \tau_c
\label{piNrel}\\ T^\pm &=& \overline{u}(p_2) \left[D^\pm(\nu,t) +
\frac{i}{2m_N} \sigma^{\mu\nu} q_{2\mu} q_{1\nu} B^\pm(\nu,t) \right]
u(p_1) \nonumber
\end{eqnarray}
with
\begin{eqnarray}
\begin{array}{llll}
s=& (p_1+q_1)^2 \; \; , & t=& (q_1-q_2)^2 \; \; , \\ u=& (p_1-q_2)^2 \; \;
, & \nu=& \displaystyle \frac{s-u}{4 m_N} \;\; .
\end{array} 
\end{eqnarray}
With the choice of invariant amplitudes $D^\pm$, $B^\pm$, the
low--energy expansion is straightforward: to determine the scattering
amplitude to $O(p^n)$, one has to calculate $D^\pm$ to $O(p^n)$ and
$B^\pm$ to $O(p^{n-2})$.

In the framework of CHPT, the first systematic calculation of 
pion--nucleon scattering was performed by Gasser et al. (1988).
In heavy baryon CHPT, the pion--nucleon scattering amplitude is not directly
obtained in the relativistic form (\ref{piNrel}) but rather as (Moj\v
zi\v s 1998)
\begin{equation}
\overline{u}(p_2) P_v^+ \left[\alpha^\pm + i \varepsilon^{\mu\nu\rho\sigma}
q_{1\mu} q_{2\nu}v_\rho S_\sigma \beta^\pm \right] P_v^+ u(p_1)~.
\label{piNH}
\end{equation}
The amplitudes $\alpha^\pm$, $\beta^\pm$ depend on the choice
of the velocity $v$. A natural and convenient
choice is the initial nucleon rest frame with $v=p_1/m_N$. In this
frame, the relativistic amplitudes can be read off directly from
Table \ref{tab:INRF}:
\begin{eqnarray}
D^\pm &=& \alpha^\pm + \displaystyle \frac{\nu t}{4 m_N} \beta^\pm
\label{eq:INRF}\\ B^\pm &=& - m_N \left(1 - \displaystyle \frac
{t}{4m^2_N} \right) \beta^\pm ~.\nonumber
\end{eqnarray}
Also the amplitudes $D^\pm$, $B^\pm$ in (\ref{eq:INRF}) will depend 
on the chosen
frame. However, as discussed before, they are guaranteed to be 
Lorentz invariant up to terms of at least $O(p^{n+1})$ if the
amplitude (\ref{piNH}) has been calculated to $O(p^n)$. 

From Eq.~(\ref{eq:DsN}) one finds that tree--level diagrams with
$D=1,2,3$ and one--loop diagrams with $D=3$ need to be calculated.
After proper renormalization, including the nonstandard nucleon
wave function renormalization, the final amplitudes 
depend on the kinematical variables $\nu,t,m_N,M_\pi$, on
the lowest--order LECs $F_\pi, g_A$, on four constants of the $p^2$
Lagrangian and on five combinations of LECs of $O(p^3)$.

The invariant amplitudes $D^\pm$, $B^\pm$ can be projected onto 
partial--wave amplitudes $f_{l\pm}^\pm(s)$. 
Threshold parameters are defined as in Eq.~(\ref{eq:effr}):
\begin{equation}
\Re e\;f_{l\pm}^\pm(s)=q^{2l}\{a_{l\pm}^\pm +q^2 b_{l\pm}^\pm +O(q^4)\}~.
\label{eq:effrpiN}
\end{equation}
To confront the chiral amplitude with experiment, \cite{MM98} has
compared 16 of these threshold parameters with the corresponding
values extrapolated from experimental data on the basis of the 
Karlsruhe--Helsinki phase--shift analysis (Koch and Pietarinen 1980).

Six of the threshold parameters ($D$ and $F$ waves) turn out to be 
independent of the low--energy constants of $O(p^2)$ and $O(p^3)$. 
The results are shown in Table \ref{tab:thresh1} and compared with 
\cite{KP80}. 

\renewcommand{\arraystretch}{1.5}
\begin{table}[t]
\caption{Comparison of two D--wave and four F--wave
threshold parameters up to the first, second and third order
(the two columns differ by higher--order terms) with 
(extrapolated) experimental values (Koch and Pietarinen 1980).
The theoretical values are based on the calculation of \protect\cite{MM98}.
Units are appropriate powers of GeV$^{-1}$.}
\label{tab:thresh1}
\begin{center}
 \begin{tabular}{|c|c|c|c|c|c|} \hline
& & & & & \\
     & \mbox{  } $O(p)$\mbox{  }  &\mbox{  }  $O(p^2)$ \mbox{  } &
   \mbox{  }   $O(p^3)$\mbox{  } & HBCHPT $O(p^3)$ &\mbox{    }
exp.\mbox{    } \\ 
& & & & & \\ \hline
\mbox{  }    $a^+_{2+} $ \mbox{  } &  $0 $ &  $-48 $  & $-35$ &  $ -36 $      
&  $  -36 \pm 7 $ \\ \hline
    $a^-_{2+} $  &  $0 $ &  $48 $  & $56$ &  $ 56 $      
&  $ 64 \pm 3 $ \\ \hline
    $a^+_{3+} $  &  $0 $ &  $0 $  & $226$ &  $ 280 $      
&   \mbox{    }$ 440 \pm 140$ \mbox{    }  \\ \hline
    $a^+_{3-} $  &  $0 $  &  $14 $  & $26$ &  $ 31$      
&  $160 \pm 120$ \\ \hline
    $a^-_{3+} $  &  $0 $      &  $0 $  & $-158$ & $-210 $      
&  $-260 \pm 20\ $ \\ \hline
    $a^-_{3-} $  &  $0 $      &  $-14 $  & $65$ &  $ 57 $   
&  $ 100 \pm 20 $ \\ \hline
  \end{tabular}
\end{center}
\end{table}

The main conclusion from Table \ref{tab:thresh1} is a definite
improvement seen at $O(p^3)$. Since there are no low--energy
constants involved (except, of course, $M_\pi$, $F_\pi$, 
$m_N$ and $g_A$), this is clear evidence for the relevance of loop
effects. The numbers shown in Table \ref{tab:thresh1} are based on the
calculation of \cite{MM98}, but essentially the same results were
obtained by \cite{BKM97}.

The altogether nine LECs beyond leading order were then fitted by
\cite{MM98} to the ten remaining threshold parameters, the $\pi
N$ $\sigma$--term and the Goldberger--Treiman discrepancy. Referring 
to \cite{MM98} for the details, let me summarize the main results:
\begin{itemize}
\item
The fit is quite satisfactory although the fitted value of the 
$\sigma$--term tends to be larger than the canonical value (Gasser et
al. 1991).
\item
In many cases, the corrections of $O(p^3)$ are sizable and definitely
bigger than what naive chiral order--of--magnitude estimates would
suggest. 
\item
The fitted values of the four LECs of $O(p^2)$ agree very well with an 
independent analysis of \cite{BKM97}. Moreover, those authors have
shown that the specific values can be understood on the basis of
resonance exchange (baryons and mesons). It seems that
the LECs of $O(p^2)$ in the pion--nucleon
Lagrangian are under good control, both numerically and conceptually.
\item
The LECs of $O(p^3)$ are of ``natural'' magnitude but more work
is needed here.
\end{itemize}

\begin{figure}[t]
\vspace*{5.5cm}
\epsfxsize=1.3cm \epsfbox[350 0 400 100]{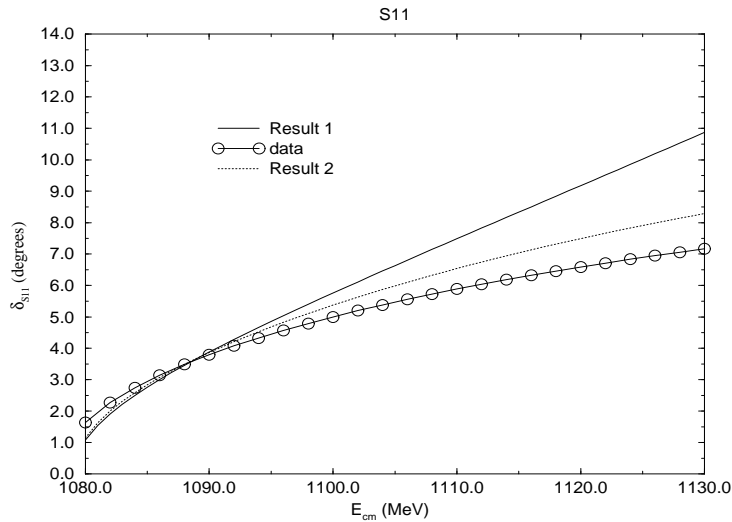}
\caption{$S_{11}$ phase shift from Datta and Pakvasa (1997).
Solid line: tree--level model with $\Delta$ and $N^*$
exchange; dotted line: complete $O(p^3)$ amplitude of Moj\v zi\v s 
(1998). Circles represent the phase shifts extracted from
fits to the $\pi N$ scattering data.}
\label{fig:s11}
\end{figure}

Using the results of \cite{MM98}, Datta and Pakvasa (1997) have
also calculated $\pi N$ phase shifts near threshold\footnote{After 
the School, a new calculation of Fettes et al. (1998) appeared where both 
threshold parameters and phase shifts are considered.}. Again, a
clear improvement over tree--level calculations can be seen in
most cases. As an example, I reproduce their results for the
$S_{11}$ phase shift in Fig.~\ref{fig:s11}.

The main conclusions for the present status of elastic $\pi N$ 
scattering are:
\begin{enumerate}
\item
The results of the first complete analysis (Moj\v zi\v s 1998)
to $O(p^3)$ in the low--energy expansion are very encouraging.
\item
Effects of $O(p^4)$ (still $L \le 1$) need to be included to check 
the stability of the expansion. 
\end{enumerate}

\subsection{Nucleon--Nucleon Interaction}

When Weinberg (1990, 1991) investigated the
nucleon--nucleon interaction within the chiral framework, he
pointed out an obvious clash between the chiral
expansion and the existence of nuclear binding. Unlike for the
meson--meson interaction that becomes arbitrarily small for
small enough momenta (and meson masses),
the perturbative expansion in the $NN$--system must break
down already at low energies. Therefore, the chiral dimension
defined in (\ref{eq:DEB}) cannot have the same interpretation as
for mesonic interactions or for single--nucleon processes.

\begin{figure}[t]
\vspace*{1cm}
\centerline{\epsfig{file=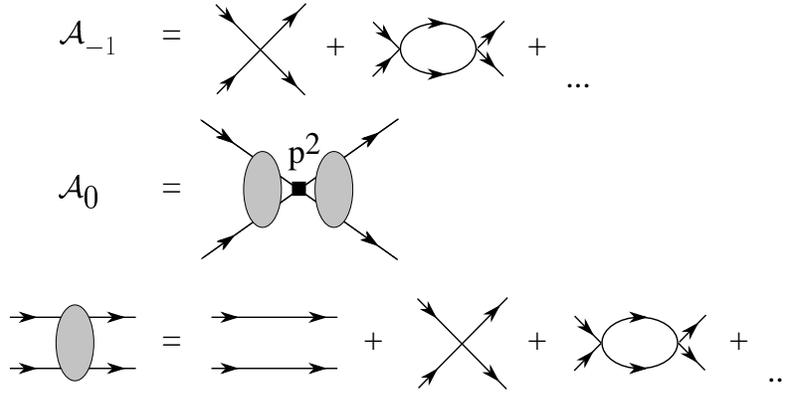,height=6cm}}
\caption{Feynman graphs contributing to the leading amplitudes for 
$^1S_0$ nucleon--nucleon scattering (from Kaplan et al. 1998).}
\label{fig:KSW1}
\end{figure}

In heavy baryon CHPT, the problem manifests itself through a
seeming infrared divergence associated with the massless
propagator of the ``light'' field $N_v$. To make the point, we
neglect pions for the time being and consider the lowest--order
four--nucleon coupling without derivatives ($n=0$ and $n_B=4$ in
the notation of Eq.~(\ref{eq:DEB})). The vertex is characterized
by the tree diagram in the first line of Fig.~\ref{fig:KSW1}. If
we now calculate the chiral dimension of the one--loop diagram
(second diagram in the first line of the figure) according to 
(\ref{eq:DEB}) we find
\begin{equation} 
D=2 L +2 - \displaystyle\frac{E_B}{2} =2 ~.
\end{equation} 
However, this result is misleading because the diagram is actually 
infrared divergent with the propagator (\ref{eq:Nvprop}).
Of course, this is an artifact of the approximation
made since nucleons are everything else but massless.
The way out is to include higher--order corrections in the
nucleon propagator. The leading correction is due to 
$\cL_{\pi N}^{(2)}$ in (\ref{eq:LpiN}). The kinetic
terms to this order are 
\begin{eqnarray}
\cL_{\rm kin} &=& \overline{N_v}\left(iv \cdot \nabla + \frac{1}{2m}
[(v \cdot \nabla)^2 - \nabla^2]\right) N_v \\
&=& \overline{N_v}\left(i \partial_0 + \frac{1}{2m}
\vec{\partial}^2\right) N_v \nonumber
\end{eqnarray}
where the last expression applies for $v=(1,0,0,0)$,
which now denotes the center--of--mass
system. The corresponding propagator in this frame is 
\begin{equation} 
\displaystyle\frac{i}{k^0 - \displaystyle\frac{\vec{k}^2}{2m}
+i\varepsilon}~.
\end{equation} 

Following Kaplan et al. (1998), we now specialize to $NN$ scattering 
in the $^1S_0$ channel and denote the incoming momenta as
\begin{equation} 
p_{1,2}=\left(\frac{E}{2}, \pm \vec{p}\right)~, \qquad
E=\frac{p^2}{m}+\dots
\end{equation} 
neglecting higher orders in the expression for the cms--energy $E$.
Including higher orders in derivatives and quark masses, Kaplan
et al. (1998) write the general tree amplitude (in $d$ dimensions) 
for nucleon--nucleon scattering in the $^1S_0$ channel as
\begin{equation}
A_{\rm tree}= - \left(\frac{\mu}{2}\right)^{4-d} \sum_{n\ge 0}
C_{2n}(\mu) p^{2n}=: - \left(\frac{\mu}{2}\right)^{4-d} C(p^2,\mu)~.
\end{equation}
The relevance of the subtraction scale $\mu$
will soon become clear. For a general vertex $C_{2n}$ of chiral
dimension $2n$, the loop diagram considered before (second
diagram in Fig.~\ref{fig:KSW1}) is easily evaluated
(Kaplan et al. 1998) in dimensional regularization:
\begin{eqnarray}
I_n &=& -i \left(\frac{\mu}{2}\right)^{4-d} \int \displaystyle
\frac{d^dk}{(2\pi)^d} \vec{k}^{2n}\displaystyle\frac{i}
{\frac{E}{2}+k^0-\frac{\vec{k}^2}{2m}+i\varepsilon}
\displaystyle\frac{i}{\frac{E}{2}-k^0-\frac{\vec{k}^2}{2m}
+i\varepsilon} \nonumber \\
&=& -m (mE)^n (-mE-i\varepsilon)^{\frac{d-3}{2}}
\Gamma\left(\frac{3-d}{2}\right)
\displaystyle\frac{(\frac{\mu}{2})^{4-d}}{(4\pi)^{\frac{d-1}{2}}}~.
\end{eqnarray}
The seeming infrared divergence of before now manifests itself as 
a divergence for $m\to\infty$. The diagram is actually
finite for $d=4$ and clearly of $O(p^{2n+1})$ invalidating the general 
formula for the chiral dimension that gave $D=2$ for $n=0$.

Kaplan et al. (1998) make the point that the diagram would be
divergent in $d=3$ dimensions with
\begin{equation} 
I_n \simeq \displaystyle\frac{m(mE)^n \mu}{4\pi(d-3)}
\end{equation} 
near $d=3$.
Although this would not seem to have any great physical significance at
first sight, Kaplan et al. (1998) suggest to subtract nevertheless
the pole at $d=3$ that actually corresponds to a linear
ultraviolet divergence in a cutoff regularization. This unconventional
subtraction procedure is in line with the observation of
other authors (e.g., Lepage 1997; Richardson et
al. 1997; Beane et al. 1998) that standard dimensional regularization
is not well adapted to the problem at hand. 

The one--loop amplitude with the subtraction prescription of Kaplan et al.
(1998) is then
\begin{equation}
I_n = - (mE)^n \frac{m}{4\pi}(\mu + ip)~.
\label{eq:In}
\end{equation}
Anticipating the following discussion, we now iterate the
one--loop diagram and sum the resulting bubble chains to
arrive at the final amplitude (Kaplan et al. 1998)
\begin{equation}
A = \displaystyle\frac{-C(p^2,\mu)}{1+\frac{m}{4\pi}(\mu + ip)
C(p^2,\mu)}~.
\label{eq:iterate}
\end{equation}
This amplitude is related to the phase shift as
\begin{equation}
e^{2i\delta} - 1 = \displaystyle\frac{ipm}{2\pi}A
\end{equation}
or, with the effective range approximation for $S$--waves
in terms of scattering length $a$ and effective range
$r_0$, 
\begin{eqnarray}
p \cot \delta &=& ip + \displaystyle\frac{4\pi}{mA}=
- \displaystyle\frac{4\pi}{mC(p^2,\mu)}-\mu \label{eq:NNeffr}\\
&=& - \frac{1}{a} + \frac{1}{2}r_0 p^2 + O(p^4) ~. \nonumber
\end{eqnarray}
Note that the (traditional) definition of the scattering length 
used here has the opposite sign compared to (\ref{eq:effr}) for
$\pi\pi$ scattering. With the relations (\ref{eq:NNeffr}), the
coefficients $C_{2n}$ can be expressed in terms of $a,r_0,\dots$:
\begin{eqnarray}
C_0(\mu) = \frac{4\pi}{m} \displaystyle\frac{1}{-\mu+1/a} & \qquad
&C_2(\mu)=\frac{2\pi r_0}{m}\left(\displaystyle\frac{1}{-\mu+1/a}
\right)^2 \label{eq:C2n}~.
\end{eqnarray}

It is known from potential scattering (e.g., Goldberger and Watson 1964) 
that $r_0$ and the higher--order coefficients in the effective
range approximation are bounded by the range of the interaction.
This also applies to $NN$ scattering in the 
$^1S_0$ channel: $r_0\simeq 2.7 \rm{~fm} \simeq 2/M_\pi$. On the other
hand, the scattering length is sensitive to states near zero
binding energy (e.g., Luke and Manohar 1997) and may be much bigger 
than the interaction range. Therefore, Kaplan et al. 
(1998) distinguish two scenarios. 
\begin{itemize}
\item Normal--size scattering length\\
In this case, also the scattering length is governed by the
range of the interaction. The simplest choice
$\mu=0$ (minimal subtraction) leads to expansion coefficients $C_{2n}$ 
in (\ref{eq:C2n}) in accordance with chiral dimensional analysis. 
This corresponds to the usual chiral expansion as in the meson or 
in the single--nucleon sector.\\
\item Large scattering length \\
In the $^1S_0$ channel of $NN$ scattering, the scattering length
is much larger than the interaction range (the situation
is similar in the deuteron channel)
\begin{equation} 
a=-23.714 \pm 0.013 \mbox{~fm} \simeq - 16/M_\pi~.
\end{equation} 
With the same choice $\mu=0$ as before, the coefficients
$C_{2n}$ are unnaturally large leading to big cancellations
between different orders. Kaplan et al. (1998) therefore suggest
to use instead $\mu=O(M_\pi)$ which leads to $C_{2n}$ of natural
chiral magnitudes.
\end{itemize}

The choice $\mu=O(M_\pi)$ immediately explains why we
have to sum the iterated loop diagrams that led to amplitude
$A$ in (\ref{eq:iterate}). Let us consider such a bubble chain graph
with coefficients $C_{2n}$ at each four--nucleon vertex. From 
(\ref{eq:C2n}) and the obvious generalization to higher--order
coefficients, one obtains $C_{2n}=O(p^{-n-1})$. Altogether, this
implies a factor $C_{2n} p^{2n} = O(p^{n-1})$ at each vertex. On the 
other hand, each loop produces a factor of order $mp/4\pi$ as can be 
seen from Eq.~(\ref{eq:In}). As a consequence,  
only the chain graphs with $C_0$ at each vertex have to be resummed
because all such diagrams are of the same order $p^{-1}$. All
other vertices can be treated perturbatively in the usual way.

The chiral expansion of the scattering amplitude (everything
still in the $^1S_0$ channel) for $\mu=O(p)$ then takes the 
form (Kaplan et al. 1998)
\begin{equation} 
A=A_{-1}+A_0+A_1+\dots
\end{equation} 
\begin{eqnarray}
A_{-1}=\displaystyle\frac{-C_0}{[1+\frac{m}{4\pi}(\mu+ip)C_0]}
& \qquad & 
A_{0}=\displaystyle\frac{-C_2 p^2}{[1+\frac{m}{4\pi}
(\mu+ip)C_0]^2}~.
\end{eqnarray}
This is also shown pictorially in Fig.~\ref{fig:KSW1}.

So far, pions have been neglected. Inclusion of pions leaves
$A_{-1}$ unchanged but modifies $A_0, A_1, \dots$. Altogether,
to next--to--leading order, $O(p^0)$, the amplitude for $NN$
scattering in the $^1S_0$ channel depends on three parameters: 
$C_0(M_\pi)$, $C_2(M_\pi)$, $D_2(M_\pi)$. Kaplan et al. (1998)
fit these three parameters to
the $^1S_0$ phase shift and obtain remarkable agreement with
the experimental phase shift all the way up to $p=300$ MeV. They
also apply an analogous procedure to the $^3S_1$ -- $^3D_1$
channels (deuteron).

After many attempts during the past years, a
systematic low--energy expansion of nucleon--nucleon scattering
seems now under control. This is an important step towards
unifying the treatment of hadronic interactions at low energies
on the basis of chiral symmetry.

\section*{Acknowledgements} 
I want to thank Willi Plessas and the members of his Organizing
Committee for all their efforts to continue the successful
tradition of the Schladming Winter School. Helpful discussions and
email exchange with J\"urg Gasser, Harald Grosse, Eduardo de Rafael
and Jan Stern are gratefully acknowledged.


\begin{thebibliography}
\bibitem{}{Ad94}{Adeva et al. 1994}
Adeva, B. et al. (DIRAC) (1994): 
Proposal to the SPSLC: Lifetime measurement of $\pi^+\pi^-$
atoms to test low energy QCD predictions,  CERN/SPSLC/P 284, Dec. 1994
\bibitem{}{AL96}{Anisovich and Leutwyler (1996)}
Anisovich, A.V., Leutwyler, H. (1996): Phys. Lett. B375, 335
\bibitem{}{Aoki98}{Aoki et al. 1998}
Aoki, S. et al. (CP-PACS) (1998):
Nucl. Phys. Proc. Suppl. 60A, 14
\bibitem{}{BF95}{Baillargeon and Franzini (1995)}
Baillargeon, M., Franzini, P.J. (1995): in \cite{MPP95} 
\bibitem{}{BC80}{Banks and Casher (1980)}
Banks, T., Casher, A. (1980): Nucl. Phys. B169, 103
\bibitem{}{BU96}{Baur and Urech (1996)}
Baur, R., Urech, R. (1996): Phys. Rev. D53, 6552
\bibitem{}{BCP98}{Beane et al. (1998)}
Beane, S.R., Cohen, T.D., Phillips, D.R. (1998): Nucl. Phys. A632, 445 
\bibitem{}{BGS94}{Bellucci et al. (1994)}
Bellucci, S., Gasser, J., Sainio, M.E. (1994):
Nucl. Phys. B423, 80; B431, 413 (E)
\bibitem{}{BKKM92}{Bernard et al. (1992)}
Bernard, V., Kaiser, N., Kambor, J., Mei\ss ner, U.-G. (1992):
Nucl. Phys. B388, 315
\bibitem{}{BKM95}{Bernard et al. (1995)}
Bernard, V., Kaiser, N., Mei\ss ner, U.-G. (1995): Int. J. Mod.
Phys. E4, 193
\bibitem{}{BKM97}{Bernard et al. (1997)}
Bernard, V., Kaiser, N., Mei\ss ner, U.-G. (1997): Nucl. Phys. A615,
483
\bibitem{}{BH95}{Bernstein and Holstein (1995)}
Bernstein, A.M., Holstein, B.R., Eds. (1995): {\it Chiral Dynamics:
Theory and Experiment}, Proc. of the Workshop at MIT, Cambridge,
July 1994 (Springer--Verlag, Berlin)
\bibitem{}{BIJ93}{Bijnens (1993)}
Bijnens, J. (1993): Phys. Lett. B306, 343
\bibitem{}{BBR}{Bijnens et al. (1993)}
Bijnens, J., Bruno, C., de Rafael, E. (1993): Nucl. Phys. B390, 501
\bibitem{}{BCG}{Bijnens et al. (1994)}
Bijnens, J., Colangelo, G., Gasser, J. (1994): Nucl. Phys. B427, 427
\bibitem{}{BEG95}{Bijnens et al. (1995)}
Bijnens, J., Ecker, G., Gasser, J. (1995): in \cite{MPP95}
\bibitem{}{BIJ96}{Bijnens (1996)}
Bijnens, J. (1996): Phys. Reports 265, 369
\bibitem{}{BCEGS1}{Bijnens et al. (1996)}
Bijnens, J., Colangelo, G., Ecker, G., Gasser, J., Sainio, M.E. (1996):
Phys. Lett. B374, 210
\bibitem{}{BP97}{Bijnens and Prades (1997)}
Bijnens, J., Prades, J. (1997): Nucl. Phys. B490, 239
\bibitem{}{BT97}{Bijnens and Talavera (1997)}
Bijnens, J., Talavera, P. (1997): Nucl. Phys. B489, 387
\bibitem{}{BCEGS2}{Bijnens et al. (1997)}
Bijnens, J., Colangelo, G., Ecker, G., Gasser, J., Sainio, M.E. (1997):
Nucl. Phys. B508, 263
\bibitem{}{BCT}{Bijnens et al. (1998a)}
Bijnens, J., Colangelo, G., Talavera, P. (1998a): The vector and
scalar form factors of the pion to two loops, hep-ph/9805389
\bibitem{}{BCE}{Bijnens et al. (1998b)}
Bijnens, J., Colangelo, G., Ecker, G. (1998b): in preparation
\bibitem{}{Bur96}{B\"urgi (1996)}
B\"urgi, U. (1996): Phys. Lett. B377, 147; Nucl. Phys. B479, 392
\bibitem{}{CCWZ69}{Callan et al. (1969)}
Callan, C.G., Coleman, S., Wess, J., Zumino, B. (1969): Phys.
Rev. 177, 2247
\bibitem{}{CDG76}{Callan et al. (1976)}
Callan, C.G., Dashen, R.F., Gross, D.J. (1976): Phys. Lett. 36B,
334
\bibitem{}{col97}{Colangelo (1997)}
Colangelo, G. (1997): Phys. Lett. B395, 289
\bibitem{}{col98}{Colangelo et al. (1998)}
Colangelo, G., Gasser, J., Leutwyler, H., Wanders, G. (1998):
in preparation
\bibitem{}{CWZ69}{Coleman et al. (1969)}
Coleman, S., Wess, J., Zumino, B. (1969): Phys. Rev. 177, 2239
\bibitem{}{CG82}{Coleman and Grossman (1982)}
Coleman, S., Grossmann, B. (1982): Nucl. Phys. B203, 205
\bibitem{}{CO84}{Collins (1984)}
Collins, J.C. (1984): {\it Renormalization} (Cambridge Univ. Press,
Cambridge)
\bibitem{}{CR77}{Crewther (1977)}
Crewther, R.J. (1977): Phys. Lett. 70B, 349
\bibitem{}{DA69}{Dashen (1969)}
Dashen, R. (1969): Phys. Rev. 183, 1245
\bibitem{}{DAP97}{Datta and Pakvasa (1997)}
Datta, A., Pakvasa, S. (1997): Phys. Rev. D56, 4322
\bibitem{}{DHW93}{Donoghue et al. (1993)}
Donoghue, J.F., Holstein, B.R., Wyler, D. (1993): Phys. Rev. D47, 2089
\bibitem{}{DP97}{Donoghue and P\'erez (1997)}
Donoghue, J.F., P\'erez, A.F. (1997): Phys. Rev. D55, 7075
\bibitem{}{DN98}{Dosch and Narison (1998)}
Dosch, H.G., Narison, S. (1998): Phys. Lett. B417, 173
\bibitem{}{threshexp}{Dumbrajs et al. (1983)}
Dumbrajs, O. et al. (1983): Nucl. Phys. B216, 277
\bibitem{}{EGPR89}{Ecker et al. (1989)}
Ecker, G., Gasser, J., Pich, A., de Rafael, E. (1989):
Nucl. Phys. B321, 311
\bibitem{}{EG94}{Ecker (1994)}
Ecker, G. (1994): Phys. Lett. B336, 508
\bibitem{}{EG95}{Ecker (1995)}
Ecker, G. (1995): Prog. Part. Nucl. Phys. 35, 1
\bibitem{}{EM96}{Ecker and Moj\v zi\v s (1996)}
Ecker, G., Moj\v zi\v s, M. (1996): Phys. Lett. B365, 312
\bibitem{}{EM97}{Ecker and Moj\v zi\v s (1997)}
Ecker, G., Moj\v zi\v s, M. (1997): Phys. Lett. B410, 266
\bibitem{}{EG97}{Ecker (1997)}
Ecker, G. (1997): Pion--pion and pion--nucleon interactions in chiral
perturbation theory, hep-ph/9710560, Contribution to the Workshop on
Chiral Dynamics 1997, Mainz, Sept. 1997, to appear in the Proceedings
\bibitem{}{fs96}{Fearing and Scherer (1996)}
Fearing, H.W., Scherer, S. (1996): Phys. Rev. D53, 315
\bibitem{}{FMS98}{Fettes et al. (1998)}
Fettes, N., Mei\ss ner, U.-G., Steininger, S. (1998): Pion--nucleon
scattering in chiral perturbation theory I: isospin--symmetric case, 
hep-ph/9803266
\bibitem{}{FSBY81}{Frishman et al. (1981)}
Frishman, Y., Schwimmer, A., Banks, T., Yankielowicz, S. (1981): 
Nucl. Phys. B177, 157
\bibitem{}{FSS90}{Fuchs et al. (1990)}
Fuchs, N.H., Sazdjian, H., Stern, J. (1990): Phys. Lett. B238, 380
\bibitem{}{FSS91}{Fuchs et al. (1991)}
Fuchs, N.H., Sazdjian, H., Stern, J. (1991): Phys. Lett. B269, 183
\bibitem{}{glplb}{Gasser and Leutwyler (1983)}
Gasser, J., Leutwyler, H. (1983): Phys. Lett. B125, 325
\bibitem{}{glann}{Gasser and Leutwyler (1984)}
Gasser, J., Leutwyler, H. (1984): Ann. Phys. (N.Y.) 158, 142
\bibitem{}{glnp1}{Gasser and Leutwyler (1985)}
Gasser, J., Leutwyler, H. (1985): Nucl. Phys. B250, 465
\bibitem{}{GSS88}{Gasser et al. (1988)}
Gasser, J., Sainio, M.E., \v Svarc, A. (1988): Nucl. Phys. B307, 779
\bibitem{}{GLS91}{Gasser et al. (1991)}
Gasser, J., Leutwyler, H., Sainio, M.E. (1991): Phys. Lett. B253, 
252, 260
\bibitem{}{GM57}{Gell-Mann (1957)}
Gell-Mann, M. (1957): Phys. Rev. 106, 1296
\bibitem{}{GMOR68}{Gell-Mann et al. (1968)}
Gell-Mann, M., Oakes, R.J., Renner, B. (1968): Phys. Rev. 175, 2195
\bibitem{}{GGRT8}{Gim\'enez et al. (1998)}
Gim\'enez, V., Giusti, L., Rapuano, F., Talevi, M. (1988):
Lattice quark masses: a nonperturbative measurement, hep-lat/9801028
\bibitem{}{GW64}{Goldberger and Watson (1964)}
Goldberger, M.L., Watson, K.M. (1964): {\it Collision Theory} (Wiley,
New York)
\bibitem{}{GO61}{Goldstone (1961)}
Goldstone, J. (1961): Nuovo Cimento 19, 154
\bibitem{}{GK95}{Golowich and Kambor (1995)}
Golowich, E., Kambor, J. (1995): Nucl. Phys. B447, 373
\bibitem{}{GK97}{Golowich and Kambor (1997)}
Golowich, E., Kambor, J. (1997): Two--loop analysis of axialvector
current propagators in chiral perturbation theory, hep-ph/9707341
\bibitem{}{GO97}{Golterman (1997)}
Golterman, M. (1997): Connections between lattice gauge theory and chiral
perturbation theory, hep-ph/9710468, Contribution to the Workshop on
Chiral Dynamics 1997, Mainz, Sept. 1997, to appear in the Proceedings
\bibitem{}{hoe83}{H\"ohler (1983)}
H\"ohler, G. (1983): in Landolt--B\"ornstein, vol. 9 b2,
Ed. H. Schopper (Springer, Berlin)
\bibitem{}{TH76}{'t Hooft (1976)}
't Hooft, G. (1976): Phys. Rev. Lett. 37, 8
\bibitem{}{TH80}{'t Hooft (1980)}
't Hooft, G. (1980): in {\it Recent Developments in Gauge Theories},
G. 't Hooft et al., Eds. (Plenum Press, New York)
\bibitem{}{Jamin98}{Jamin (1998)}
Jamin, M. (1998): Nucl. Phys. Proc. Suppl. 64, 250
\bibitem{}{JM91}{Jenkins and Manohar (1991)}
Jenkins, E., Manohar, A.V. (1991): Phys. Lett. B255, 558
\bibitem{}{KWW96}{Kambor et al. (1996)}
Kambor, J., Wiesendanger, C., Wyler, D. (1996): Nucl. Phys. B465, 215
\bibitem{}{KM86}{Kaplan and Manohar (1986)}
Kaplan, D.B., Manohar, A.V. (1986): Phys. Rev. Lett. 56, 1994
\bibitem{}{KSW98}{Kaplan et al. (1998)}
Kaplan, D.B., Savage, M.J., Wise, M.B. (1998): A new expansion
for nucleon--nucleon interactions, nucl-th/9801034; Two--nucleon
systems from effective field theory, nucl-th/9802075
\bibitem{}{KV88}{Kazakov (1988)}
Kazakov, D. (1988): Theor. Math. Phys. 75, 440 
\bibitem{}{KSSF93}{Knecht et al. (1993)}
Knecht, M., Sazdjian, H., Stern, J., Fuchs, N.H. (1993):
Phys. Lett. B313, 229
\bibitem{}{KMSF95}{Knecht et al. (1995)}
Knecht, M., Moussallam, B., Stern, J., Fuchs, N.H. (1995):
Nucl. Phys. B457, 513
\bibitem{}{KMSF96}{Knecht et al. (1996)}
Knecht, M., Moussallam, B., Stern, J., Fuchs, N.H. (1996):
Nucl. Phys. B471, 445
\bibitem{}{KU97}{Knecht and Urech (1997)}
Knecht, M., Urech, R. (1997): Virtual photons in low--energy $\pi\pi$
scattering, hep-ph/9709348
\bibitem{}{KR97}{Knecht and de Rafael (1997)}
Knecht, M., de Rafael, E. (1997): Patterns of spontaneous chiral
symmetry breaking in the large--$N_c$ limit of QCD--like theories,
hep-ph/9712457
\bibitem{}{KP80}{Koch and Pietarinen (1980)}
Koch, R., Pietarinen, E. (1980): Nucl. Phys. A336, 331
\bibitem{}{leefranz}{Lee--Franzini (1997)}
Lee--Franzini, J. (KLOE) (1997): Contribution to the Workshop on
Chiral Dynamics 1997, Mainz, Sept. 1997, to appear in the Proceedings
\bibitem{}{LRT97}{Lellouch et al. (1997)}
Lellouch, L., de Rafael, E., Taron, J. (1997): Phys. Lett. B414, 195
\bibitem{}{LE97}{Lepage (1997)}
Lepage, G.P. (1997): How to renormalize the Schr\"odinger equation,
Lectures given at the 8th Jorge Andre Swieca Summer School, Sao Paolo,
Brazil, Feb. 1997
\bibitem{}{hl90}{Leutwyler (1990)}
Leutwyler, H. (1990): Nucl. Phys. B337, 108
\bibitem{}{LS92}{Leutwyler and Smilga (1992)}
Leutwyler, H., Smilga, A. (1992): Phys. Rev. D46, 5607
\bibitem{}{hlann}{Leutwyler (1994)}
Leutwyler, H. (1994): Ann. Phys. (N.Y.) 235, 165
\bibitem{}{hl96a}{Leutwyler (1996a)}
Leutwyler, H. (1996a): Phys. Lett. B374, 181
\bibitem{}{hl96b}{Leutwyler (1996b)}
Leutwyler, H. (1996b): Phys. Lett. B378, 313
\bibitem{}{hl96c}{Leutwyler (1996c)}
Leutwyler, H. (1996c): Light quark masses, hep-ph/9609467,
Carg\`ese Lectures 1996
\bibitem{}{hl97}{Leutwyler (1997)}
Leutwyler, H. (1997): Probing the quark condensate by means of $\pi\pi$
scattering, hep-ph/9709406, Proceedings of the DA$\Phi$CE Workshop, 
Frascati, Nov. 1996
\bibitem{}{Lowe}{Lowe (1997)}
Lowe, J. (BNL-E865) (1997): Contribution to the Workshop on
Chiral Dynamics 1997, Mainz, Sept. 1997, to appear in the Proceedings
\bibitem{}{LM92}{Luke and Manohar (1992)}
Luke, M., Manohar, A.V. (1992): Phys. Lett. B286, 348
\bibitem{}{LM97}{Luke and Manohar (1997)}
Luke, M., Manohar, A.V. (1997): Phys. Rev. D55, 4129
\bibitem{}{LU97}{L\"uscher (1997)}
L\"uscher, M. (1997): Theoretical advances in lattice QCD, 
hep-ph/9711205, Talk given at the 18th Int. Symposium on 
Lepton--Photon Interactions, Hamburg
\bibitem{}{MPP95}{Maiani et al. (1995)}
Maiani, L., Pancheri, G., Paver, N., Eds. (1995):
{\it The Second DA$\Phi$NE Physics Handbook} (INFN, Frascati)
\bibitem{}{MG84}{Manohar and Georgi (1984)}
Manohar, A.V., Georgi, H. (1984): Nucl. Phys. B234, 189
\bibitem{}{MMS97}{Mei\ss ner et al. (1997)}
Mei\ss ner, U.-G., M\"uller, G., Steininger, S. (1997): Phys. Lett. B406,
154; B407, 454 (E)
\bibitem{}{MM98}{Moj\v zi\v s (1998)}
Moj\v zi\v s, M. (1998): European Phys. Journal 2, 181
\bibitem{}{MOU97}{Moussallam (1997)}
Moussallam, B. (1997): Nucl. Phys. B504, 381
\bibitem{}{NJL61}{Nambu and Jona-Lasinio (1961)}
Nambu, Y., Jona-Lasinio, G. (1961): Phys. Rev. 122, 345
\bibitem{}{OK62}{Okubo (1962)}
Okubo, S. (1957): Prog. Theor. Phys. 27, 949
\bibitem{}{PS97}{Post and Schilcher (1997)}
Post, P., Schilcher, K. (1997): Phys. Rev. Lett. 79, 4088
\bibitem{}{Prades98}{Prades (1998)}
Prades, J. (1998): Nucl. Phys. Proc. Suppl. 64, 253
\bibitem{}{EdR98}{de Rafael (1998)}
de Rafael, E. (1998): An introduction to sum rules in QCD, 
hep-ph/9802448, Lectures delivered at Les Houches Summer School 1997
\bibitem{}{RBM97}{Richardson et al. (1997)}
Richardson, K.G., Birse, M.C., McGovern, J.A. (1997): Renormalization
and power counting in effective field theories for nucleon--nucleon
scattering, hep-ph/9708435 
\bibitem{}{ross}{Rosselet et al. (1977)}
Rosselet, L. et al. (1977): Phys. Rev. D15, 574
\bibitem{}{Schacher}{Schacher (1997)}
Schacher, J. (DIRAC) (1997): Contribution to the Workshop on
Chiral Dynamics 1997, Mainz, Sept. 1997, to appear in the Proceedings
\bibitem{}{SH97}{Sharpe 1997}
Sharpe, S.R. (1997): Nucl. Phys. Proc. Suppl. 53, 181
\bibitem{}{SVZ79}{Shifman et al. (1979)}
Shifman, M.A., Vainshtein, A.I., Zakharov, V.I. (1979):
Nucl. Phys. B147, 385, 447
\bibitem{}{SSF93}{Stern et al. (1993)}
Stern, J., Sazdjian, H., Fuchs, N.H. (1993): Phys. Rev. D47, 3814
\bibitem{}{stern97}{Stern (1997)}
Stern, J. (1997): Light quark masses and condensates in QCD, 
hep-ph/9712438, Contribution to the Workshop on
Chiral Dynamics 1997, Mainz, Sept. 1997, to appear in the Proceedings
\bibitem{}{stern98}{Stern (1998)}
Stern, J. (1998): Two alternatives of spontaneous chiral
symmetry breaking in QCD, hep-ph/9801282, submitted to Phys.
Rev. Letters
\bibitem{}{UR68}{Urban (1968)}
Urban, P., Ed. (1968): {\it Particles, Currents, Symmetries}, Proc. of 
the 7. Int. Universit\"atswochen f\"ur Kernphysik, Schladming, 1968, 
Acta Phys. Austriaca, Suppl. V (Springer--Verlag, Wien, New York)
\bibitem{}{VW84}{Vafa and Witten (1984)}
Vafa, C., Witten, E. (1984): Nucl. Phys. B234, 173; Comm. Math.
Phys. 95, 257
\bibitem{}{WA98}{Walcher (1998)}
Walcher, T., Ed. (1998):  Proceedings of the Workshop on
Chiral Dynamics, Mainz, Sept. 1997, in preparation
\bibitem{}{wein66}{Weinberg (1966)}
Weinberg, S. (1966): Phys. Rev. Lett. 17, 616
\bibitem{}{wein67}{Weinberg (1967)}
Weinberg, S. (1967): Phys. Rev. Lett. 18, 507
\bibitem{}{wein77}{Weinberg (1977)}
Weinberg, S. (1977): in {\it A Festschrift for I.I. Rabi}, L. Motz, 
Ed. (New York Academy of Sciences, N.Y.)
\bibitem{}{wein79}{Weinberg (1979)}
Weinberg, S. (1979): Physica 96A, 327
\bibitem{}{wein90}{Weinberg (1990)}
Weinberg, S. (1990): Phys. Lett. B251, 288
\bibitem{}{wein91}{Weinberg (1991)}
Weinberg, S. (1991): Nucl. Phys. B363, 3
\bibitem{}{WZ71}{Wess and Zumino (1971)}
Wess, J., Zumino, B. (1971): Phys. Lett. 37B, 95
\bibitem{}{WI83}{Witten (1983)}
Witten, E. (1983): Nucl. Phys. B223, 422

\end{thebibliography}
\end{document}